\documentclass[sigconf, screen]{acmart}

\AtBeginDocument{%
  \providecommand\BibTeX{{%
    \normalfont B\kern-0.5em{\scshape i\kern-0.25em b}\kern-0.8em\TeX}}}

\setcopyright{acmcopyright}
\acmPrice{15.00}
\acmDOI{10.1145/3540250.3549111}
\acmYear{2022}
\copyrightyear{2022}
\acmSubmissionID{fse22main-p295-p}
\acmISBN{978-1-4503-9413-0/22/11}
\acmConference[ESEC/FSE '22]{Proceedings of the 30th ACM Joint European Software Engineering Conference and Symposium on the Foundations of Software Engineering}{November 14--18, 2022}{Singapore, Singapore}
\acmBooktitle{Proceedings of the 30th ACM Joint European Software Engineering Conference and Symposium on the Foundations of Software Engineering (ESEC/FSE '22), November 14--18, 2022, Singapore, Singapore}

\usepackage[ruled,vlined]{algorithm2e}
\usepackage{multirow}
\usepackage{graphicx}
\usepackage{subfig}
\usepackage{framed}
\usepackage{url}
\usepackage{makecell}
\usepackage{appendix}
\usepackage{enumerate}
\usepackage{enumitem}
\usepackage{tcolorbox}
\usepackage{mdframed}
\newcounter{finding}
\newenvironment{finding}
{
    \refstepcounter{finding}
	\begin{mdframed}[
    	nobreak=true,
    	linecolor=black,
    	roundcorner=12pt,
    	backgroundcolor=gray!05,
    	linewidth=0.5pt,
    	leftmargin=0.5em,
    	rightmargin=0.5em,
    	topline=true,
    	bottomline=true,
    	frametitlerule=true,
    	frametitlebackgroundcolor=gray!30,
    	frametitlerulecolor=gray,
    	frametitle=Common Practice \arabic{finding},
    	frametitleaboveskip=0.3em,
    	frametitlebelowskip=0.35em,
    	skipabove=12pt
	]
}
{
    \end{mdframed}
    \vspace{1em}
}

\newcommand\todo[1]{}

\newcommand\change[1]{\textcolor{black}{#1}}
\newcommand\jz[1]{}
\newcommand\lou[1]{}

\newcommand{\surveyNumber}{100 }

\newcommand{\ADSTestingNumber}{42 }

\newcommand{\AllNumber}{117 }
\newcommand{\ADSAndTestingNumber}{102 }

\begin{document}

\title{Testing of Autonomous Driving Systems: Where Are We and Where Should We Go?}


\author{Guannan Lou}
\affiliation{%
  \institution{Macquarie University}
  \city{Sydney}
  \state{NSW}
  \country{Australia}}
\email{guannan.lou@mq.edu.au}

\author{Yao Deng}
\affiliation{%
  \institution{Macquarie University}
  \city{Sydney}
  \state{NSW}
  \country{Australia}}
\email{yao.deng@hdr.mq.edu.au}

\author{Xi Zheng}
\authornote{Corresponding authors: Xi Zheng, Mengshi Zhang.}
\affiliation{%
  \institution{Macquarie University}
  \city{Sydney}
  \state{NSW}
  \country{Australia}}
\email{james.zheng@mq.edu.au}

\author{Mengshi Zhang}
\authornotemark[1]
\affiliation{%
  \institution{Meta}
  \city{Menlo Park}
  \state{CA}
  \country{USA}}
\email{mengshizhang@fb.com}

\author{Tianyi Zhang}
\affiliation{%
  \institution{Purdue University}
  \city{West Lafayette}
  \state{IN}
  \country{USA}}
\email{tianyi@purdue.edu}


\begin{abstract}

Autonomous driving has shown great potential to reform modern transportation. Yet its reliability and safety have drawn a lot of attention and concerns. Compared with traditional software systems, autonomous driving systems (ADSs) often use deep neural networks in tandem with logic-based modules. This new paradigm poses unique challenges for software testing. Despite the recent development of new ADS testing techniques, it is not clear to what extent those techniques have addressed the needs of ADS practitioners. To fill this gap, we present the first comprehensive study to identify the current practices and needs of ADS testing. We conducted semi-structured interviews with developers from 10 autonomous driving companies and surveyed 100 developers who have worked on autonomous driving systems. A systematic analysis of the interview and survey data revealed 7 common practices and 4 emerging needs of autonomous driving testing. Through a comprehensive literature review, we developed a taxonomy of existing ADS testing techniques and analyzed the gap between ADS research and practitioners' needs. Finally, we proposed several future directions for SE researchers, such as developing test reduction techniques to accelerate simulation-based ADS testing.
\end{abstract}


\maketitle

\section{Introduction}
Autonomous driving has been making great strides towards reality in recent years. In 2017, Waymo launched the trail of fully autonomous ride-hailing services in California \cite{gibbs2017google}. More recently, Tesla released the beta version of its Full Self-Driving (FSD) software, which has been installed on at least 60K Tesla vehicles~\cite{tesla}. However, given the traffic accidents caused by autonomous vehicles
~\cite{ADSAccident, CADMV, favaro2018autonomous} 
, there is still a long way to ensure the robustness and reliability of autonomous driving systems (ADSs). Similar to traditional software systems, autonomous driving companies also adopt testing as the main quality assurance mechanism for ADSs. Several leading companies such as Waymo and Tesla have their own fleets to perform extensive on-road testing. Furthermore, simulation testing environments such as Carla~\cite{carla} are widely adopted to test various driving scenarios or extreme conditions that cannot be easily replicated in reality.

In recent years, the Software Engineering (SE) community has also proposed many testing techniques to improve the safety and reliability of ADSs~
\cite{gambi2019asfault, tang2021systematic, li2020av, ishikawa2020testing, arcaini2021targeting, borg2021digital, calo2020generating, abdessalem2018vision, gambi2019automatically, kluck2019genetic, mullins2018adaptive, lu2021search, luo2021targeting, abdessalem2018testing, gladisch2019experience, zhang2018deeproad, tian2018deeptest, valle2021metamorphic, han2020metamorphic, chandrasekaran2021combinatorial, gambi2019generating, kluck2018using, tao2019industrial, gambi2019reconstructing, li2020ontology, zhou2020deepbillboard, kim2020reducing, wolschke2017observation, garcia2020comprehensive, liu2021analysis, jahangirova2021quality, knauss2017software, 9282713, hu2021disclosing, leudet2019ailivesim, haq2020comparing, masuda2017software, fritzsch2021experiences, du2021towards,king2019automated,salay2019safety, zhao2019assessing}.
For example,  DeepRoad and DeepTest \cite{zhang2018deeproad, tian2018deeptest} leverage metamorphic testing to test ADSs under extreme weather conditions. AC3R~\cite{gambi2019generating} generates critical driving scenarios (e.g. collisions) in a simulation environment based on crash reports. These methods have shown promising results for end-to-end driving models. However, with the rapid evolution of autonomous driving technologies, industrial ADSs have become much more sophisticated these days, using multiple perception models in tandem with logic-based control and planning modules. Currently, there is a lack of comprehensive understanding of the emerging needs of ADS testing and to what extent existing ADS testing techniques meet the needs of ADS practitioners. To bridge this gap, we adopted a mixed methods research design~\cite{johnson2004mixed} 
with a combination of qualitative interview study, large-scale survey, and literature review to investigate the following three research questions:
\vspace{-1mm}
\begin{itemize}[leftmargin=*]
    \item \textbf{RQ1}. \textit{What are the industrial practices of ADS testing?}
    \item \textbf{RQ2.} \textit{What are the emerging needs of ADS testing?} 
    \item \textbf{RQ3.} \textit{To what extent  existing ADS testing techniques address the industrial needs?} 
\end{itemize}
\vspace{-1mm}

We first interviewed developers from 10 autonomous driving companies to collect qualitative responses on ADS testing practices and needs. Based on the insights from the interview study, we developed a  survey and solicited quantitative responses from a boarder audience. Specifically, we sent out 1978 surveys to ADS developers and received 100 complete and valid responses. Through comprehensive data analysis and triangulation, we summarized seven common practices in ADS development and testing, some of which have not been accounted by existing testing techniques. For example, ADS practitioners adopt more system-level testing metrics such as consistency and latency when testing an ADS, rather than just using model accuracy. 
Furthermore, we identified four common needs of ADS testing: (1) {\em identifying possible corner cases and unexpected driving scenarios}, (2) {\em speeding up ADS testing}, (3) {\em tool support for constructing complex driving scenarios}, and (4) {\em tool support for data labeling}.

We further conducted literature review on ADS testing methods proposed by the SE community. We manually went through \AllNumber papers that mentions autonomous driving or related keywords in 28 SE conferences and identified \ADSTestingNumber papers specifically about ADS testing. We categorized them and developed a taxonomy of different kinds of research on ADS testing. We identified four gaps between existing research and industrial needs. 
For instance, existing ADS testing methods only consider simple image transformations. There is a lack of support for identifying richer and more complex traffic scenarios that ADS developers really need in practice. Based on these gaps, we proposed several future directions. For example, SE researchers can leverage test selection and prioritization techniques to speed up ADS testing (Need 2). Specifically, to account for the multi-module architecture of ADSs, test selection methods can be formulated as a multi-objective optimization problem to maximize multiple coverage metrics for 
different modules (e.g., neural coverage for a perception model and statement coverage for a logic-based control module), rather than a single model.

\textbf{Paper organization.} Section~\ref{sec:related} discusses related work in ADS testing. Section~\ref{sec:Methodology} illustrates how we conduct interviews and surveys. Section~\ref{sec:interview} provides current practice of ADS testing in industry. Section~\ref{sec:needs} summarizes needs of industry in ADS testing. Section~\ref{sec:future} discusses the existing SE solutions to these needs and future research directions. Section~\ref{sec:validity} discusses the 
threats to validity in this study. Section~\ref{sec:conclusion} concludes 
this work.
\section{Related Work} \label{sec:related}
There are several literature reviews on testing and verification of machine learning (ML) models~\cite{zhang2020machine, xiang2018verification, Dey_2021multilayered, borg2018safely}. The most comprehensive literature review is by Zhang et al.~\cite{zhang2020machine}. They analyzed 114 papers related to machine learning testing and provided an overview of various testing properties, components, and workflows of ML models. In addition, Zhang et al.~identified several challenges of testing ML models, such as how to generate natural test inputs. Compared with these studies, our study focuses on ADS testing. In addition to a literature review, we also interviewed and surveyed ADS practitioners to understand the common practices and needs of ADS testing. We found that the unique characteristics of ADSs (e.g., the multi-module architecture) along with the special testing needs pose new challenges and opportunities compared with ordinary ML models, e.g., leveraging multi-objective search in test selection for multi-module ADSs.

Two recent studies analyzed the codebase and bugs in autono-mous driving systems~\cite{peng2020first, garcia2020comprehensive}. Peng et~al.~\cite{peng2020first} conducted a case study of Baidu Apollo~\cite{Apollo} and summarized the architecture and individual modules in the Apollo system. They found that Apollo lacked adequate testing at the system level.
Garcia et al.~\cite{garcia2020comprehensive} analyzed 16,851 commits and 499 issues in Apollo and Autoware~\cite{autoware}. They classified these issues and summarized their symptoms and root causes. There are also several empirical studies on bugs in ML applications~\cite{zhang2019empirical, zhang2018empirical, thung2012empirical, sun2017empirical, amershi2019software}. For example, Zhang et al.~\cite{zhang2018empirical} analyzed 175 software bugs in ML applications built by TensorFlow. 

The most related work to our study includes several literature reviews on autonomous driving testing~\cite{huang2016autonomous, koopman2016challenges, masuda2017software}. Huang et al.~\cite{huang2016autonomous} reviewed testing and verification methods from the Intelligent Vehicle (IV) community. They summarized not only testing methods for individual modules in the software stack, but also hardware-in-the-loop testing methods for hardware components and integrated testing for the entire vehicle. Compared with Huang et al., we focus on testing methods from the software perspective. Our literature review summarized recent advances in ADS testing from the SE community. Please refer to Section~\ref{sec:future} for a detailed summary and discussion of these software testing techniques. Koopman and Wagner proposed five challenges of testing autonomous vehicles based on the V model for autonomous vehicles~\cite{koopman2016challenges}. Masuda described the software architecture of autonomous vehicle simulations and discussed several software testing challenges of such simulations~\cite{masuda2017software}. Compared with them, our discussion is anchored upon common practices and needs of ADS testing from interviews and online surveys with ADS practitioners. 
\section{Methodology} \label{sec:Methodology}
\begin{figure*}[tb]
    \centering
    \includegraphics[width = 0.7\textwidth]{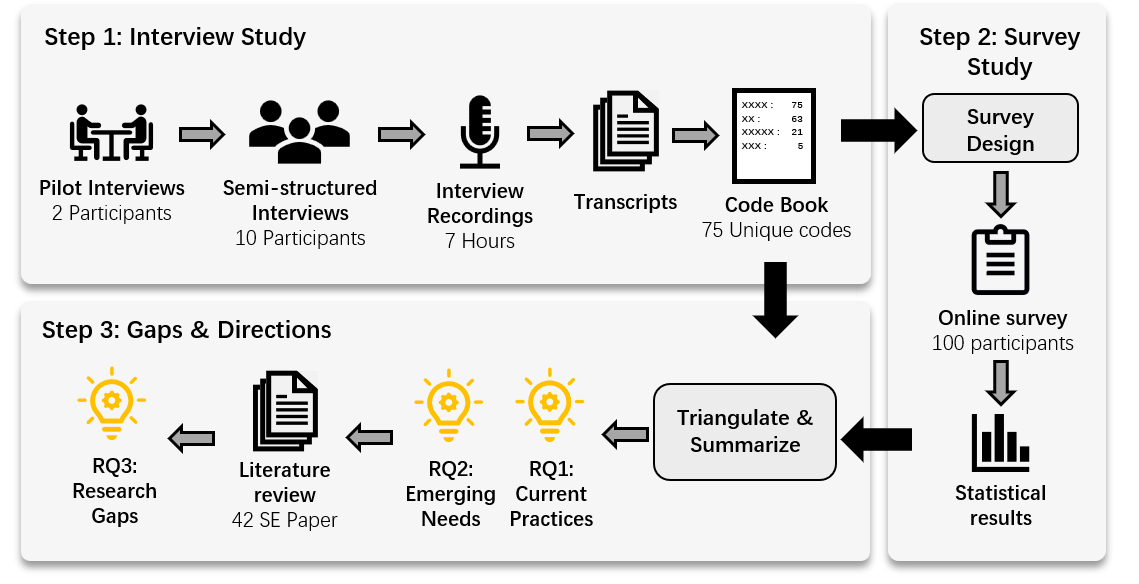}
    \caption{Research methodology}
    \label{fig:research_methdology}
\end{figure*}

Following \cite{DBLP:books/sp/08/SSS2008}, Our empirical study contains 3 steps, as shown in Figure~\ref{fig:research_methdology}. 
First, we conducted semi-structured interviews with 10 developers from 10 different autonomous driving companies.  Based on the findings from these interviews, we further conducted a large-scale survey with 100 ADS practitioners to quantitatively validate our findings from the interviews. We then summarized and triangulated the findings from the surveys and the interviews. Third, we conducted an literature review of SE papers related to ADS testing. We categorized those papers and developed a taxonomy of ADS testing techniques. By comparing the taxonomy and the emerging needs of ADS practitioners, we identified the research gaps. 


\subsection{Interviews}

\textbf{Interview protocol.} We designed an interview guide\footnote{The complete interview guide is publicly available here: \url{https://bit.ly/3DKia0C}} for the semi-structured interviews. The interview began with a short introduction of our study. Then, we asked high-level questions about: 1) the background and expertise of interviewees, such as the current job role and how long they have been working on ADSs, 2) the current practices, methodologies, and tools they used for ADS testing, and 3) the challenges and difficulties they faced in ADS testing. We first conducted 2 pilot interviews, based on which we further refined the interview guide. Then, we interviewed 10 developers from 10 different autonomous driving companies. Each interview took between 30 minutes and an hour. The interviews were recorded with permission of participants and then transcribed for analysis.


    {\begin{table}[t]
        \centering
        
        \caption{Interview participant background}
                \vspace{-5pt}
        \label{tab:participant info}
        \begingroup
        \renewcommand\arraystretch{1.1}
        \scalebox{0.9}{
        \begin{tabular}{lllll}
            \hline
             & \multicolumn{1}{c}{\textsf{Experience}} & \multicolumn{1}{c}{\textsf{Responsibilities}} \\ 
            \hline
            P1  & 2 years   & Recognition algorithm design and testing \\
            P2  & 2 years   & Simulation testing and real-world testing \\
            P3  & 6 years   & Passenger car development \\
            P4  & 1.5 years    & Perception system development \\
            P5  & 10 years  & ADS testing \\
            P6  & 3 years   & ADS testing \\
            P7  & 1.5 years    & Perception system testing \\
            P8  & 5 years    & ADS developing and testing \\
            P9  & 1.5 years  & Recognition algorithm design and testing \\
            P10 & 1.5 years    & ADS testing, Simulation testing\\ 
            \hline
        \end{tabular}}
    \endgroup

    \end{table}}

\textbf{Participants.} We recruited 10 software developers from 10 different autonomous driving companies based on our personal network, industrial collaboration, and social media. As shown in Table~\ref{tab:participant info}, these participants had at least one and a half years of experience (3.4 years on average) in ADS development and testing. 
In addition, their responsibilities also covered every aspect of ADS testing, including modular testing, simulation-based testing,  and real-world testing. 

\textbf{Analysis.} We transcribed a total of 7 hours of interview recordings to text using an audio transcription tool called iflyrec\footnote{\url{https://www.iflyrec.com/}}. We manually corrected errors in the transcripts. The first two authors conducted inductive thematic analysis \cite{braun2006using}\footnote{The maxqda codebook is publicly available here: \url{https://bit.ly/38K9Csk}} using a qualitative data analysis software called MAXQDA\footnote{\url{https://www.maxqda.com/}}. Specifically, they first independently labeled the transcripts to extract relevant or insightful responses from participants and summarize them into short descriptive texts, which are called codes. Then they met each other, compared their codes, and discussed any inconsistencies. They continuously refined the codes during the discussion. Then they gro-uped related codes into themes. Results of the thematic analysis were regularly reported and discussed with the whole research team. The final inter-rater agreement ratio of the first two authors is 0.85, measured by Cohen's Kappa~\cite{viera2005understanding}. 

\subsection{Survey}
Since the findings in the interview study are only based on responses from 10 participants, we further conducted a large-scale survey to validate the interview findings and solicit more feedback from a broader population of ADS practitioners. 

\textbf{Survey design.} 
We designed a survey\footnote{The complete survey form is publicly available here: \url{https://bit.ly/2WNVV93}} with 
33 questions in 5 sections, including \textit{background}, \textit{autonomous driving system}, \textit{testing practices and challenges 
in ADS}, \textit{testing needs in ADS}, and \textit{follow up}. The \textit{background} section asked about participants' background, expertise, and their current roles in the company or organization. 
The \textit{autonomous driving system} section asked about the ADSs participants have worked on.
The \textit{testing practices and challenges on ADS} section asked about three types of ADS testing identified from the previous interviews, including unit testing, simulation testing, and on-road testing. For each type of testing, we designed multiple-choice questions based on the findings from the interview study. The options in a multiple-choice question were derived from responses of the interview study. Participants can also select ``Others'' to supplement alternative answers, or select ``I don't know'' to indicate they have no insights in the particular question. For each type of testing, we also included an open-ended question in the end to solicit additional feedback on what participants would like to have or improve on. The \textit{testing needs in ADS} section listed 
four common needs identified in the interview study. We designed a linear scale question for each need and asked participants to provide a numeric response to 
indicate the importance of each need in a 7-point likert scale, from ``unimportant at all'' (1) to ``very important'' (7). The \textit{follow up} section asked participants their contact information and whether they would like to participate in any follow-up study.

\textbf{Participants.} We recruited survey participants in three ways. First, we sent out surveys to the autonomous driving companies where our interview participants came from. Second, we used the APIs provided by Twitter and LinkedIn to scrape the contact information (if any) of developers 
with profile associated with autonomous driving.
Finally, we searched for popular autonomous driving software repositories on Github such as Apollo~\cite{Apollo} and DeepDrive~\cite{DeepDrive}. Then, we manually identified the public email address (if any) of those developers 
contributed to those repositories.

In total, we sent out 1978 surveys and received 114 responses, with a response rate of 5\%. We discarded 14 survey responses since they are not complete. In the end, we collected \surveyNumber survey responses that are complete for data analysis.\footnote{The survey result is publicly available here: \url{https://bit.ly/38K9Csk}}
90\% of survey participants were male, 8\% were female, and 2\% did not disclose their gender identity. 54\% of them worked in research institutes, 36\% worked in technology companies, 7\% of participants worked in traditional vehicle manufactures, and 3\% of participants were self-employed. Regarding their job roles, 33\% of participants were in R\&D positions, 14\% were in management positions, and 24\% were engineers, including safety engineers, software engineers and perception engineers. 40\% of participants had two to three years of working experience in ADS, 29\% had worked in ADS for more than three years, and 31\% had less than two years of experience.

\textbf{Analysis.} 
For each multiple-choice question, we plotted the choices made by survey participants in a histogram, such as Figure~\ref{fig:test cases real}. Specifically, if a survey participant supplemented an alternative answer that was not identified in our interview study, we first checked whether it was similar to an existing choice. If not, we considered it as a unique answer and created a short, descriptive label for it. We then merged all the alternative answers with the same label and plotted their distributions in the histogram as well.
For linear-scale questions, we calculated the total number and percentage of each option, and used the percentage to illustrate survey participants’ evaluation of the importance of industrial needs, such as Figure~\ref{Fig:need}.
For open-ended questions, the first two authors used MAXQDA to code each answers individually, classified these answers according to sections in this work, and discussed with each other to refine codes and classifications.

\subsection{Literature Review}
\label{sec:literature_method}


\change{We created the literature review protocol including research question, literature search strategy, literature selection criteria, literature selection procedures, data extraction strategy, and synthesis of the extracted data, following the methodology described in \cite{keele2007guidelines} to guide the literature review process. Our research question of literature review is to investigate whether existing research works related to ADS testing are adequate for the challenges and needs identified in our interviews and surveys. The literature search strategy is to search research papers from $28$ SE conferences, journals and workshops including ICSE, ESEC/FSE, ASE, ISSTA, ISSRE, ICST, QRS, TSE, JSS, and IST. Specifically, we used $35$ keywords to identify research papers related to autonomous driving, e.g., ``autonom-ous driving'', ``autonomous vehicle'', ``traffic scene'', ``Apollo'', etc. We built a crawler to download papers from publisher websites and found \AllNumber papers that contain at least one of the keywords in their title or abstract. The literature selection criterion is to include technical papers that propose ADS testing methods and empirical papers that discuss challenges and solutions related to ADS testing. We manually reviewed all searched papers and found that \ADSAndTestingNumber papers related to ADS, among which \ADSTestingNumber are specifically about ADS testing\footnote{The keyword list, venue list and paper list are publicly available at \url{https://bit.ly/3FLUwCz}}. We categorized these \ADSTestingNumber ADS testing papers into different categories based on their \change{research questions and solutions}. The result is summarized in Section~\ref{sec:taxonomy}.}

\section{Common Practices of ADS Testing} 
\label{sec:interview}

This section summarizes the common practices of ADS testing. Section~\ref{sec:ADStype} discusses the types of ADSs that our interview and survey participants worked on. Section \ref{sec:unit}, Section \ref{sec:realworld} and Section \ref{sec:simulation} discuss three commonly used ADS testing methods---\textit{unit testing}, \textit{real-world testing} and \textit{simulation testing}. We do not separately present the results of the interview study and the survey for two main reasons. First, since the purpose of the survey is to confirm and supplement the qualitative findings from the interview, the findings from the interview study and the survey study have a lot of overlap. Therefore, if we report the findings separately, there will be a lot of redundancy. Second, fusing quantitative evidence (such as statistics from a large-scale survey) and qualitative evidence (such as quotations from interview participants) is more convenient for readers to read and understand the results of the survey. 

\subsection{Autonomous Driving Systems under Test} \label{sec:ADStype}
    
        


    Our participants mainly work on two main types of ADSs: \textit{multi-module driving systems} and \textit{end-to-end driving models}. 
    
    \textbf{Multi-module driving systems.} The majority of interviewees (100\%) and survey participants (69\%) developed and tested multi-module driving systems.  Multi-module architectures are widely used in industry-scale driving systems, e.g., Autoware~\cite{kato2018autoware} and Apollo~\cite{Apollo}. They contain several modules for perception, prediction, planning, and control. The perception module takes a variety of sensor data as input, such as road images, point clouds, and GPS signals, to detect surrounding objects. The prediction module predicts the moving trajectories of these surrounding objects. Given the perception and prediction results, the planning module then decides on the route of the ego-vehicle. Finally, the control module converts the planned route to vehicle control commands, including braking force and steering angle. Four interview participants (P1, P3, P7, P9) elaborated that the perception and prediction modules in their systems heavily use deep neural networks, while the planning and control modules typically contain traditional logic-based programs. This is consistent with a previous study on Apollo~\cite{wang2019apolloscape}.
    
    \textbf{End-to-end driving models.} 
    31\% of survey participants said they worked on end-to-end (E2E) driving models, while none of the interviewees worked on E2E driving models.
    E2E driving models such as PilotNet~\cite{bojarski2017explaining} and OpenPilot~\cite{openpilot} treat the entire driving pipeline as a single deep learning model, from processing sensor data to generating vehicle controls. During the interview, participants pointed out that E2E driving models are less preferred in the industry, since they cannot handle complicated driving scenarios and often suffer from generalizability issues. For example, P7 said,  \textit{``Training an E2E driving model requires a large amount of data. And E2E driving models can  easily over-fit and perform worse than multi-module systems.''}

     
     \begin{finding}{} {The majority of ADS practitioners reported to work on multi-module ADSs rather than end-to-end driving models. Therefore, multi-module ADSs deserve more attention in future research.
     }\end{finding}

\subsection{Unit testing}\label{sec:unit}
80\% of interviewees and 52\% of survey participants reported that they conducted unit testing during ADS development.

\textbf{\textit{Testing target.}} 
In addition to writing unit tests for control logic, ADS developers also need to test DL models, especially those models in the perception and prediction modules. Unlike program source code, these models do not have clearly control logic or program states. ADS developers need to manually label and clip driving recordings collected from on-road or simulation testing to construct small recording segments for testing these DL models, which are referred to as unit tests in ADS development.
There interviewees (P1, P4, P9) reported that there are too many possible driving scenarios to test, and it is time-consuming to manually process driving recordings to construct test scenarios.

In addition, 
70\% of interviewees and over 68\% of survey participants said their driving systems used at least three of four types of sensors, e.g., cameras, LiDARs, radars, GPS. 
This multi-modality of sensor data makes it more difficult to generate test cases. For example, data from different sensors with different sample frequencies need to be synchronized (e.g., by timestamp). Furthermore, when transforming one type of sensor data, such as adding an object, other types of sensor data must be updated consistently. 


\begin{finding}{}
In addition to testing control logic, ADS developers also need to construct segments of driving recordings to test DL models, which take multi-modal sensor data as input, not just road images.
\end{finding}



\textbf{\textit{Test metrics.}} When measuring the performance of a model, ADS developers use metrics, including accuracy, precision, recall, Receiver Operating Characteristic (ROC), and Intersection over Union (IoU). Furthermore, 51\% of survey participants said they also use specific metrics tailored for different ADS modules. Two interview participants (P1, P9) elaborated on this---consistency is one of their metrics when testing the lane detection model. The lane detection results between the front and back frames in a video frame should be consistent. 


\begin{finding}{}
ADS practitioners not only use 
common model performance metrics, such as accuracy and IoU,
but also custom-tailored metrics such as consistency when testing perception models in ADS.
\end{finding}

\subsection{Real-world testing}\label{sec:realworld}{
    
    All interview participants and 57\% of survey participants reported that they have done real-world testing. Two types of real-world testing are 
    mentioned: {\em scenario-based testing} and {\em on-road testing}. 
    
    \begin{figure}[tb]
    \centering
    \includegraphics[width = 0.85\linewidth]{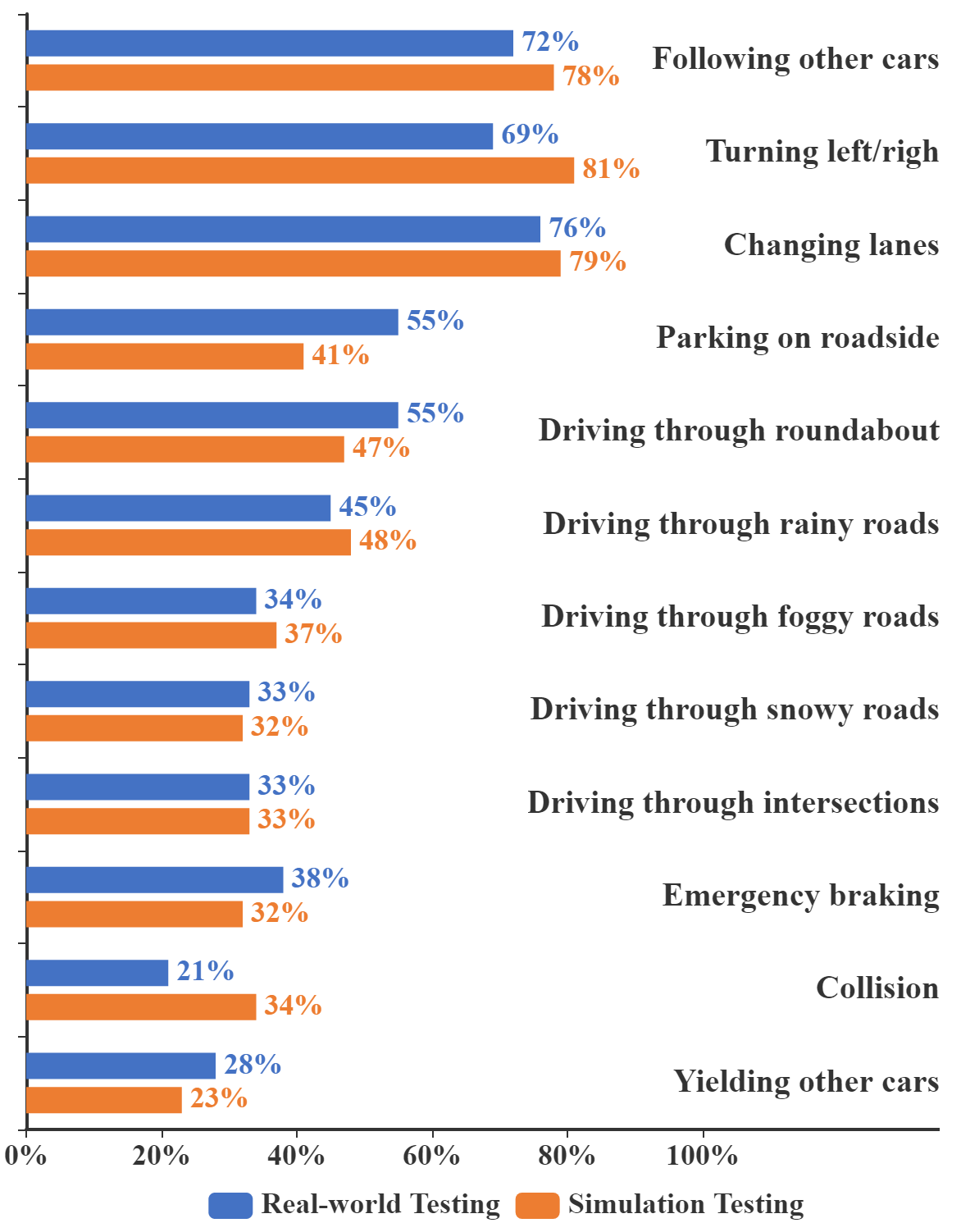}
    \caption{Common driving scenarios tested in on-road testing and simulation testing, respectively}
    \label{fig:test cases real}
    \end{figure}
 
    
    \begin{figure}[tb]
    \centering
    \includegraphics[width = 0.45\textwidth]{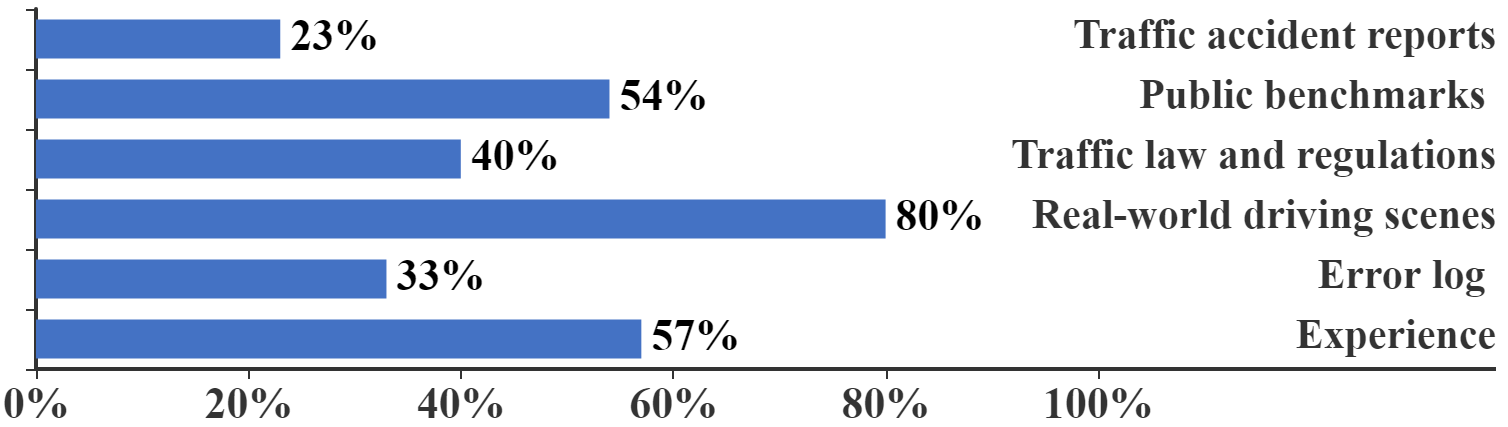}
    \caption{Sources of tested driving scenarios}
    \label{fig:case_comefrom}
    \end{figure}
    
    {\bf\em Scenario-based testing.}
    4 interview participants (P1, P2, P8, P9) and 89\% survey participants mentioned they have tested specific driving scenarios in closed automated vehicle proving grounds. Figure \ref{fig:test cases real} shows commonly tested driving scenarios, as reported in our survey. It includes common driving scenarios such as changing lanes (76\%), following other cars (72\%), and turning left or right (69\%), as well as special road sections such as intersections (33\%) and roundabouts (over 55\%). Weathers such as rainy days (44\%) and snowy days (33\%) are considered as well. Finally, dangerous driving scenarios such as emergency braking and collision are also conducted by 28\% and  21\% of survey participants, respectively. In the intervew study, P1 and P2 said they have built their own scenario database following traffic law and regulations. P8 and P9 said setting up those driving scenarios were time-consuming, which often took 2 weeks to 1 month.
    
    Figure \ref{fig:case_comefrom} shows the sources where those driving scenarios were from, as reported in the survey . 80\% and 57\% of survey participants said those scenarios were based on common real-world driving scenes and driving experiences. 54\% of survey participants said they referred to public benchmarks such as CityScapes~\cite{Cordts2016Cityscapes}, ApolloScape~\cite{wang2019apolloscape}, and Waymo open dataset~\cite{sun2020scalability}. 40\% of survey participants said they referred to traffic laws and regulations. 23\% said they referred to  traffic accident reports.
    
    
    \begin{finding}{}
    ADS practitioners design various driving scenarios based on real-world driving scenes, public benchmarks, traffic laws and regulations, and crash reports to test an ADS in the field.
    \end{finding}

    \begin{figure}[tb]
        \centering
        \includegraphics[width = 0.4\textwidth]{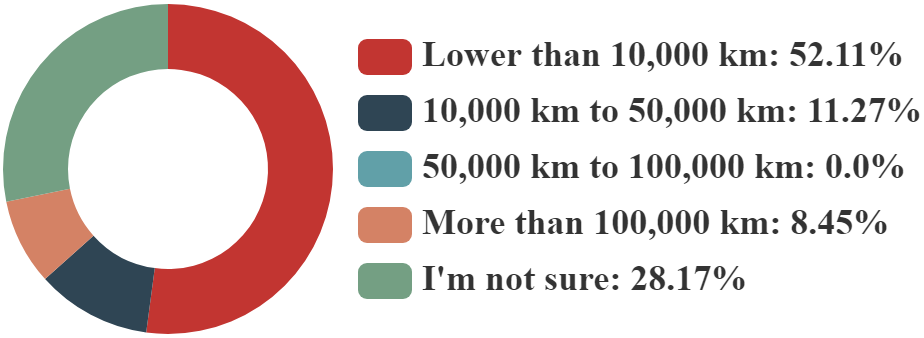}
        \caption{Length of on-road testing performed by industrial survey participants}
        \label{fig:driving}
    \end{figure}

    {\bf\em On-road testing.}\label{sec:onroad} 
    In on-road testing, autonomous vehicles are tested in public roads where various scenarios and unexpected conditions could occur. In practice, ADS companies need to carry out long-distance on-road testing to ensure the reliability of their driving systems. The number of kilometers an ADS travels without human intervention (i.e., km per disengagement) in on-road testing is used to measure the reliability of their ADS. As shown in the Disengagement from California Department of Motor Vehicles~\cite{Disengagement}, Waymo conducted over 1 million kilometers of on-road testing in 2020, and its distance for each disengagement was about 48,000 km.
    During the interview, P8 said, ``\textit{the entire on-road testing required 50,000 to 100,000 kilometers, which must include different driving scenes such as highways, country roads, and urban roads.}'' Furthermore, P9 mentioned that testers had to sit in the car and record driving data in on-road testing. 
    The collected data would be either saved in local storage or uploaded to cloud platform, some of which would be cleaned and labelled for model training and testing. While on-road testing is critical, the reality is that, due to technical and financial constraints, many ADS practitioners can only conduct on-road testing within a limited range of mileage. Figure~\ref{fig:driving} shows that only 8\% of participants conducted more than 100,000 km on-road testing, and 11\% of them conducted 10,000 to 50,000 km on-road testing. As discussed in the next section, simulation testing is often considered as a more affordable and safer testing option for the majority of ADS practitioners.
    
    
    \begin{finding}{}
    On-road testing is considered as critical to ensure ADS reliability and robustness, while only a small portion of ADS practitioners have done long-distance testing. 
    \end{finding}
    
    \begin{figure}[tb]
    \centering  
    \includegraphics[width = 0.45\textwidth]{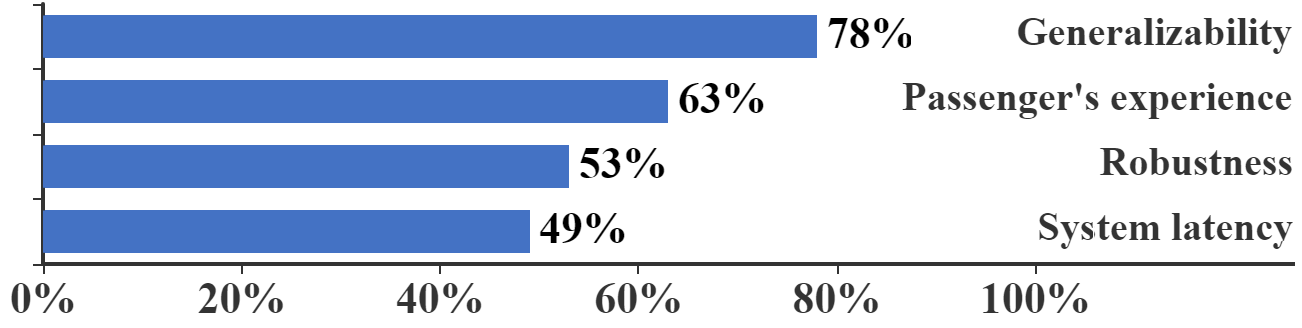}
    \caption{System-level metrics used in real-world testing}
    \label{Fig:metirc}
    \end{figure}
    
    {\bf\em Test metrics.}
    Compared with metrics used in unit testing, more system-level metrics are used in real-world testing. As shown in Figure \ref{Fig:metirc}, four system-level metrics are frequently mentioned by survey participants, including generalizability (78\%), passenger's experience (63\%), robustness (53\%), and system latency (49\%). Generalizability focuses on whether an ADS can achieve similar performance in unseen driving scenes. Passenger's experience is another important metric in real-world testing. P4 said, ``\textit{when there are many vehicles in a driving scene, an ADS should not press the brake too frequently, which may make passengers uncomfortable.}''
    Robustness assesses whether an ADS can behave normally when noises, external interference, or attacks exist.
    As an ADS is expected to make quick, continuous responses to the change of driving scenes, system latency is often used to measure the decision-making speed of an ADS.
    
    
    \begin{finding}{}
    Test metrics used in real-world testing focus more on system-level performance, including both functional and non-functional properties, rather than only model accuracy. 
    \end{finding}
}

\subsection{Simulation testing}\label{sec:simulation} 

 5 interviewees and 87\% of survey participants said they conducted ADS testing in a simulation platform such as Carla~\cite{carla}, AirSim~\cite{AirSim} and LGSVL~\cite{lgsvl}. 
 When asked why simulation testing is so widely practiced, participants mentioned two main reasons. First, 54\% of survey participants supported that modern simulation environme-nts are powerful enough to test corner cases that are costly to set up in the real world. In the interview study, P2 and P6 mentioned that for dangerous driving scenarios such as collisions, simulation testing is much more preferred due to safety concerns. As shown in Figure \ref{fig:test cases real}, more participants did collision test in simulation than in the real world. Second, P2 and P9 mentioned that simulators can replay sensor data and vehicle control commands collected from real-world testing, which can be used to test the performance of a new release of an ADS. 55\% of survey participants also support this point of view. Simulation testing is an important part of regression testing in ADS development. When developing an ADS, it is costly and time-consuming if a developer tests every commit of ADS in the real-world. Leveraging the simulation platform, industrial practitioners do not need to deploy each commit of ADS on the real vehicle and construct real-world test scenarios. 


\begin{finding}{}
Simulation testing is widely adopted as a complement for real-world testing. It is particularly adopted to test new commits as part of regression testing.
\end{finding}
    



{\bf\em Test metrics.} Test metrics used in simulation testing is similar as on-road testing, including accuracy-based metrics, generalizability, passenger's experience, and robustness. However, as the simulation platform is built as an ideal environment without hardware delay, system latency is less considered in simulation testing.

\section{Emerging Needs}\label{sec:needs}

    \begin{figure}[tb]
    \centering  
    \includegraphics[width = 0.45\textwidth]{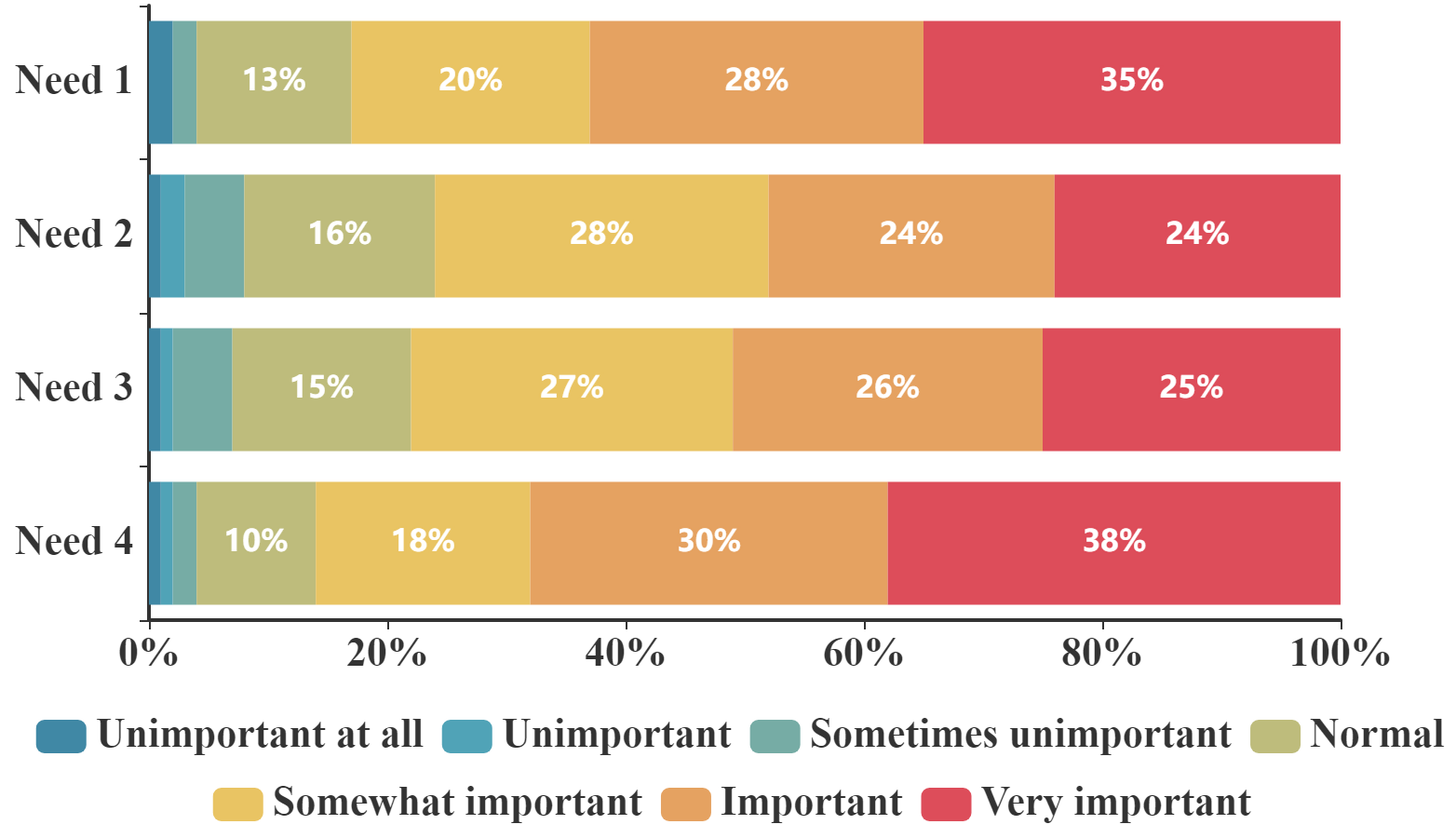}
    \caption{The importance of four common needs voted by survey participants. Details of each need are described in Section \ref{sec:needs}.}
    \label{Fig:need}

    \end{figure}

   This section summarizes four emerging needs of ADS testing identified in the interview and survey. Figure \ref{Fig:need} shows survey participants' agreement on the importance of these needs. 
   68\%, 63\%, 51\% and 48\% of survey participants considered Need 4, Need 1, Need 3, and Need 2 important or very important. \change{Especially for Need 4 and Need 1, they are regarded as very important by more than one-third of survey participants.}

    \subsection{Need 1: Identifying possible corner cases and unexpected driving scenarios} \label{sec:need1}

    As shown in Figure \ref{Fig:need}, 
    35\% and 28\% of survey participants rated this need \textit{very important} and \textit{important} respectively. While modern driving systems are tested with 
    diverse driving scenarios and extensive on-road testing, new corner cases are still often found during on-road testing. As P9 said, \textit{``During on-road testing on California highways, we found it difficult to distinguish lane lines under the sunset. This was a problem we did not expect when we perform scenario-based testing.''} Based on our conversation with ADS practitioners, the current industry practice to discover more corner cases 
    seems to be simply performing more and longer on-road testing. However, unlike large automotive companies, small companies often lack resources (e.g., vehicle fleets) and on-road testing certificates to perform large-scale on-road testing. Therefore, identifying corner cases efficiently is an urgent need for ADS practitioners.
    
     \subsection{Need 2: Speeding up ADS testing}
     \label{sec:need2}
        
        As shown in Figure \ref{Fig:need}, 24\% and 24\% of survey participants rated this need \textit{very important} and \textit{important} respectively. 
        According to some existing work~\cite{butler1991infeasibility, koopman2016challenges}, the catastrophic failure rate of an ADS should be minimized to $10^{-7}$ to $10^{-9}$ for 1 to 10 hours driving to achieve the goal of high reliability. To verify that the 
        failure rate falls within one per $10^{7}$ hours, one must conduct at least $10^{7}$-hour testing of an ADS (about 1141 vehicle-years). It is unrealistic to achieve this goal with on-road testing. Therefore, ADS practitioners often resort to simulation testing. Widely-used simulators such as CarSim~\cite{benekohal1988carsim} support acceleration that only takes one-tenth of the time compared with real-world testing. However, given the large amount of time required to achieve high reliability (i.e., $10^{7}$-hour driving test), it is still time consuming to conduct test in simulation. As P2 said, \textit{``Our original goal was to do 20,000 kilometers of testing. Later, we found that even if it accelerates the speed by 5 times, the speed of the test is still slow, compared with our expectation.''}
        Therefore, simulation acceleration is not a silver bullet to this challenge. Other methods to speed up ADS testing are needed. For example, test selection and prioritization approaches have been widely investigated in tradition software systems, which can be leveraged to accelerate ADS testing. We discuss this  in detail in Section~\ref{sec:need2_solution}. 
        
    \subsection{Need 3: Tool Support for Constructing Complex Driving Scenarios} \label{sec:need3}
    
    As shown in Figure \ref{Fig:need}, 
    25\% and 26\% of survey participants rated this need \textit{very important} and \textit{important} respectively. ADS practitioners design driving scenarios based on various sources such as real-world driving scenes and traffic accident reports (Section~\ref{sec:realworld}). After identifying driving scenarios worth testing, they need to construct corresponding test cases. 65\% of survey participants found it cumbersome to use toolkits (e.g., domain-specific languages, libraries, etc.) provided by existing simulation platforms to construct a test case of a driving scenario. Take OpenScenario~\cite{OpenScenario} as an example. It takes 258 lines of code to construct a simple driving scenario of lane cutting, not even to mention complex scenarios. The main reason is that existing simulation platforms only provide low-level APIs and domain-specific languages (DSLs) to construct driving scenarios. While such low-level API and language design ensures the flexibility and expressiveness to precisely specify arbitrary driving scenarios, it imposes significant coding effort. Similar to how Google releases Keras as a higher-level abstraction of TensorFlow, it would be beneficial to provide a higher-level abstraction of the APIs and DSLs in the simulation platforms. In addition, as mentioned by P8, "\textit{We already collected traffic accident video from internet, but it is still hard to automatically transform them into test cases in simulators.}" ADS developers wish they can get more tool support that helps them automatically or semi-automatically construct driving scenarios in a simulation environment, such as translating a natural language description of a traffic scene into the low-level code written in APIs provided by a simulation environment.
    
    \subsection{Need 4: Tool support for data labeling} \label{sec:need4}
    
    As shown in Figure \ref{Fig:need}, 38\% and 30\% of survey participants rated this need \textit{very important} and \textit{important} respectively. 
    Section~\ref{sec:realworld} has already discussed that driving data collected in on-road testing, such as point clouds and road images, are often used as test data. However, as mentioned by P8, \textit{these driving data need to be manually labelled and clipped first before they can be replayed in a simulation environment or reused to test a DL model.} 
    Two typical labels are 2D bounding boxes and semantic segmentation masks. A 2D bounding box label includes the height and width of a detected object and the type of the object. And a semantic segmentation mask requires data labelers to manually segment all objects at pixel level. Given the massive amount of driving data collected from on-road testing, the labeling effort is enormous. While there have been some assistive labeling tools in recent years~\cite{joshi2020amazon, VoTT, sharma2019labelbox}, labeling driving data still requires many manual effort. For example, Amazon SageMaker~\cite{joshi2020amazon} can automatically assign labels for object detection and image segmentation tasks. But it requires data labelers to manually go through those labels and make corrections. Several ADS companies we have interviewed said they often outsourced the data labeling task to data labeling companies or used crowdsourcing platforms such as Amazon Mechanic Turks~\cite{paolacci2014inside}. However, even for manually labeled data, label quality is still a concern, especially for safety-critical tasks like autonomous driving. According to a recent study~\cite{northcutt2021pervasive}, several well-known datasets such as ImageNet are riddled with manual labeling mistakes. Therefore, ADS practitioners wish to get more tool support to analyze, recognize, and repair labeling errors in their driving data.  
  
 \begin{figure*}[tb]
    \centering
    \includegraphics[width = 1\textwidth]{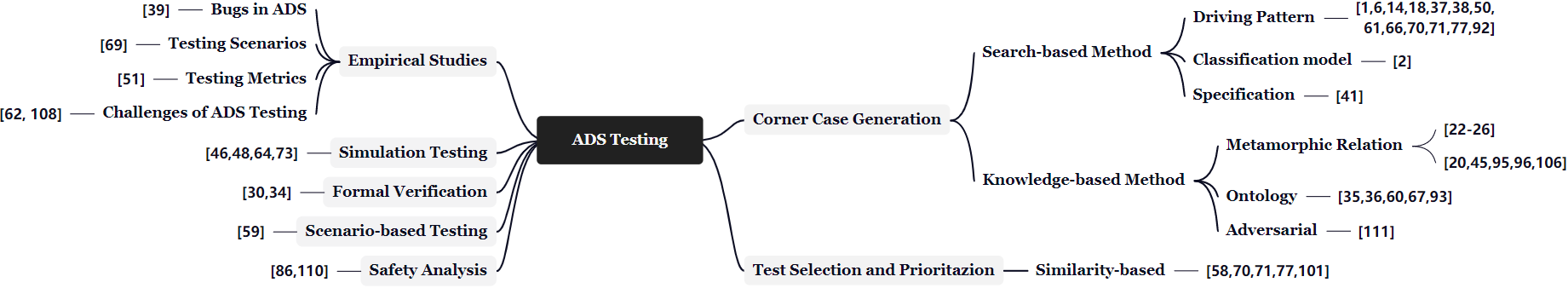}
    \caption{A taxonomy of ADS testing research.}
    \label{fig:ADS_papers}
\end{figure*}

\section{Literature Review and Research Gaps} \label{sec:future}

The previous section summarizes four emergent needs from industrial practitioners in ADS testing based on interviews and surveys. To understand to what extent state-of-the-art ADS testing techniques have addressed these needs, we surveyed research papers from software engineering conferences and journals and manually assessed them. This section summarizes our findings and future research opportunities.

\subsection{Literature Taxonomy}
\label{sec:taxonomy}
Figure~\ref{fig:ADS_papers} describes our taxonomy of ADS testing techniques, based on the \ADSTestingNumber papers identified from Section~\ref{sec:literature_method}. 

\vspace{1mm}
\noindent{\bf Corner case generation.} A variety of techniques have been proposed to generate corner cases for ADSs. We categorized them into two lines of research. One line of research is {\em knowledge-based methods}
~\cite{zhang2018deeproad, tian2018deeptest, valle2021metamorphic, han2020metamorphic, chandrasekaran2021combinatorial, gambi2019generating, kluck2018using, tao2019industrial, gambi2019reconstructing, li2020ontology, zhou2020deepbillboard}.
Knowledge-based methods generate corner cases using domain-specific ontology or metamorphic relations, which are designed based on human driving experience, traffic accident reports, traffic law and regulations. For example, DeepTest~\cite{tian2018deeptest} and DeepRoad~\cite{zhang2018deeproad} generated corner cases using  metamorphic relations from common sense that weather changes
should not affect the steering angle prediction of an ADS.
AC3R~\cite{gambi2019generating} used domain-specific ontology and natural language processing 
to extract information from police crash reports and reconstruct corner cases of crash accidents. 
DeepBillboard~\cite{zhou2020deepbillboard} generated corner cases by adding adversarial perturbation~\cite{deng2020analysis} patches on real-world billbo-ards to check whether the E2E driving model keeps the same output of the steering angle.

{\em Search-based method}~\cite{gambi2019asfault, tang2021systematic, li2020av, ishikawa2020testing, arcaini2021targeting, borg2021digital, calo2020generating, abdessalem2018vision, gambi2019automatically, kluck2019genetic, mullins2018adaptive, lu2021search, luo2021targeting, abdessalem2018testing, gladisch2019experience} is another line of research in corner case generation for ADSs. These methods aim to find a set of test parameters that introduce behavior discrepancies in an ADS, such as collisions and lane departures. 
To define a parameter search space, the majority of search-based methods leverage {\em driving patterns}~\cite{gambi2019asfault, tang2021systematic, li2020av, ishikawa2020testing, arcaini2021targeting, borg2021digital, calo2020generating, abdessalem2018vision, gambi2019automatically, kluck2019genetic, mullins2018adaptive, lu2021search, luo2021targeting}. Driving patterns are designed based on the states of an autonomous vehicle, such as its position and speed, and other vehicles and pedestrians on the road. The search process is normally guided by manually defined fitness functions such as time-to-collision and the minimal distance between the ego-vehicle and pedestrians. Optimal solutions of such fitness functions are found by applying genetic algorithms~\cite{gambi2019asfault, luo2021targeting, abdessalem2018vision, gambi2019automatically}. For example, AsFault~\cite{gambi2019asfault} used  genetic algorithm to evaluate parameters of road, like length and position, to make the ego-vehicle more error-prone on lane keeping task on generated test cases. Abdessalem et al.~\cite{abdessalem2018testing} proposed a search algorithm that combines the genetic algorithm with a decision tree classifier to guide the test case generation faster towards critical test scenarios. Gladisch et al.~used Bayesian optimization as an alternative for genetic algorithms~\cite{gladisch2019experience}.

\vspace{1mm}
\noindent{\bf Test selection and prioritization.} We identified 5 papers~\cite{mullins2018adaptive, lu2021search, luo2021targeting, kim2020reducing, wolschke2017observation} on this research direction. These methods select or prioritize test cases according to the similarity between the vector representation of each test case. 
Kim et al.~\cite{kim2020reducing} encoded a test case as an embedding vector using the output of a set of selected neuron outputs, and used the likelihood-based surprise adequacy and Mahalanobis-distance-based surprise adequacy to measure the difference between a test case with others by estimating the density of vector representations of each selected test case. Test cases with high surprise adequacy would be selected as critical test cases.


\vspace{1mm}
\noindent{\bf Empirical Studies.} We found 5 empirical studies on ADS testing. 
Knauss et al.~\cite{knauss2017software} and Zhang et al.~\cite{9282713} conducted empirical studies including focus groups and interviews to identify challenges in ADS testing. Jahangirova et al.~\cite{jahangirova2021quality} performed a systematic analysis on metrics of driving qualities, and used a mutation analysis to select 26 metrics, such as the max lateral position and the standard error of speed, as functional oracles for ADS testing. These selected oracles can kill mutants of the E2E driving model with low false alarm rate.
Liu et al.~\cite{liu2021analysis} analyzed autonomous vehicle disengagement and collision reports from the California Department of Motor Vehicles~\cite{CADMV} and found that the growth of on-road testing mileage is not accompanied by the increase of the disengagement ratio.
Garcia et al.~\cite{garcia2020comprehensive} investigated on commits and ADS bugs in Apollo~\cite{Apollo} and Autoware~\cite{autoware}, and classified found bugs into 13 root causes, such as incorrect algorithm implementations, incorrect condition logic and concurrency.



\vspace{1mm}
\noindent{\bf Simulation Testing.}
4 SE papers have discussed simulation testing.
Hu et al.~\cite{hu2021disclosing} and Masuda et al.~\cite{masuda2017software} discussed the potential issues in simulation testing, including huge input space and precision issues in the simulator.
Haq et al.~\cite{haq2020comparing} discussed the reliability of driving data generated by simulators.
Leudet et al.~\cite{leudet2019ailivesim} designed a simulator called AILiveSim, which is not only focused on autonomous vehicles, but can also be used to simulate autonomous ships and autonomous mining machines. 

\vspace{1mm}
\noindent{\bf Formal Verification.} 
Fritzsch et al.~\cite{fritzsch2021experiences} reported their experience of employing bounded model checking on a symbolic model checker NuSMV to verify relevant properties for a vehicle control system. 
Du et al.~\cite{du2021towards} used a DSL to specify relationships among driving scenario elements and used Stochastic Hybrid Automata to specify the dynamic behaviors, which can then be checked by an existing model checker UPPAAL-SMC.  The approach can be used to verify safety-related properties of a specific driving scenario.


\vspace{1mm}
\noindent{\bf Scenario-based testing.} 
King et al.~\cite{king2019automated} proposed a scenario-based testing method to test the adaptive cruise control function in ADS. The proposed method can parallel evaluate multiple systems in multiple similar but unknown test scenarios. And the prototype is 

\vspace{1mm}
\noindent{\bf Safety Analysis.}
Salay et al.~\cite{salay2019safety} introduced a classification-specific safety analysis tool based on failure mode effects analysis. The method named CFMEA is able to identify failure modes and risk levels in perception models of ADS. Similarly Zhao et al.~\cite{zhao2019assessing} used a variant of Conservative Baysian inference to avoid catastrophic failure due to overconfidence for a few inference tools used in AV.

\subsection{Gaps and Future Directions for Need 1}
\label{sec:Need1GF}
Among the four emerging needs identified in our study, the need of identifying corner cases and unexpected driving scenarios is the one that receives the most attention from the research community. As discussed in Section~\ref{sec:taxonomy}, a large number of corner case generation techniques have been proposed for ADSs. These techniques have already constituted of good solutions that we believe ADS practitioners should try out. Here we summarize several improvement opportunities for the existing techniques. First, for ontology-based or metamorphic relation-based methods, it is worth investigating the generation of more complex driving scenarios (e.g., lane merging, overtaking) beyond, for instance, affine transformations and weather conditions. 
Second, for search-based methods, it is always challenging to define a tractable search space and fitness function for realistic, complex driving scenarios. As a future direction, it is worth investigating how to elicit new safety requirements from existing crash reports, traffic rules and on-road driving recordings. Such requirements can be used to generate new fitness functions or refine existing ones to augment the search space for more mission-critical testing scenarios. 

Furthermore, as discussed in Section~\ref{sec:ADStype}, all interview participants and 69\% of survey participants are now working on multi-module ADSs. These multi-module ADSs take multiple types of sensor data as input. Yet the majority of test generation techniques only generate road image data. Therefore, it is worth investigating how to generate multi-modal sensor data for new driving scenarios.

\subsection{Gaps and Future Directions for Need 2} \label{sec:need2_solution}
Five papers on test selection and prioritization can be applied to address the need of accelerating ADS testing~\cite{mullins2018adaptive, lu2021search, kim2020reducing, wolschke2017observation, luo2021targeting}. 
These techniques can be used to remove redundant testing scenarios and prioritize driving recordings that are more likely to expose errors in an ADS. For example, when driving on a  highway, the ego-vehicle often goes straight on a specific lane at a constant speed. Therefore, there is no need to replay the entire highway driving recording but only unique driving scenarios during the recording, such as lane cutting. \todo{Check if the following writing is still true after adding the new ADS test prioritization papers} All these technical papers use similarity-based test metrics. Researchers can also investigate other kinds of  metrics, such as confidence-based metrics~\cite{feng2020deepGini, ma2021test, Li_2020PEACEPACT, wang2021prioritizing}, to select or prioritize test cases. For example, DeepGini~\cite{feng2020deepGini} applied Gini impurity on confidence to measure the likelihood of misclassification on an input, and prioritized test cases with high Gini scores. Furthermore, instead of using test metrics, we also suggest that researchers consider dynamic HD maps, which is an intermediate representation of driving scenes used by modern ADSs. HD maps incorporate the outputs of perception models and are then fed as input for the following prediction and control modules. They can be used as a holistic representation for driving scenes.

Since current ADS test prioritization techniques focus on end-to-end (E2E) driving models or only the path planning module in an ADS, there is an opportunity to develop new techniques for multi-module ADSs. To account for multi-module architectures, researchers may want to investigate how to combine test metrics for inner model behaviors, such as neuron coverage and surprise adequacy, and logic-based test metrics, such as path coverage, to capture the interaction among multiple models, and interaction between models and logic-based control modules. One possible solution is to formulate this as a multi-objective optimization problem. A technique should search for a minimal set of test cases that collectively maximize the coverage of individual ML models as well as the path or dependency coverage of an ADS. For example, a driving scene at a busy traffic intersection is likely to have higher collective coverage at the model and system levels compared with a highway scene with little traffic, since it involves multiple types of objects (e.g., pedestrians, vehicles, and traffic lights) that may trigger multiple execution paths in an ADS and also trigger multiple perception models. Given the large number of possible test case combinations, it is worth investigating how to leverage multi-objective optimization algorithms to efficiently search for an optimal test set.

\subsection{Gaps and Future Directions for Need 3}

We have only identified two relevant SE papers~\cite{gambi2019generating, du2021towards} that addressed the need of constructing complex driving scenarios. Gambi et al.~\cite{gambi2019generating} proposed an automated approach called AC3R that parses crash reports to driving scenarios in a simulation environment. Specifically, AC3R leverages NLP techniques to extract key information (e.g., traffic participants and driving actions) from crash reports and transform crash information to a Lua script to control traffic participants and generate corresponding crash scenarios. AC3R is limited to some simple crash scenarios like rear-end collision of two vehicles. In addition, as crash reports often follow rigid narrative standards, AC3R may have a difficult time translating free-form driving scene descriptions to an executable driving scene. As future work, it is worth investigating new techniques in NLP and CV to support automated construction of driving scenes from text or video data. Du et al.~\cite{du2021towards} proposed a new modeling language that describes specifications in various driving scenarios for the purpose of formal verification. While this new language is not directly used to construct new test cases, it can be adapted as a domain-specific language (DSL) for test generation. 

We also searched more broadly online and found some effort in addressing this need from other research communities. The Intelligent Vehicle and Programming Language communities has proposed several higher-level DSLs to reduce the effort of modeling objects and motions in a simulation platform~\cite{fremont2019scenic, queiroz2019geoscenario, althoff2017commonroad, bender2014lanelets}. For instance, Scenic~\cite{fremont2019scenic} 
is a probabilistic programming language which enables developers to generate complex traffic jam scenarios with just a few lines of code. These newly designed DSLs are more compact and flexible. We believe SE researchers can also contribute to this line of research. 

\subsection{Gaps and Future Directions for Need 4}


Among all four emerging needs, the need of tool support for data labeling receives the highest importance rating in the survey study. 68\% of survey participants considered this need as very important or important. We have only identified one SE paper~\cite{kim2020reducing} that attempts to address this need. Kim et al.~\cite{kim2020reducing} proposed to surprise adequacy, a simple metric to quantify how surprising an input is to a DNN, to guide the selection of new road images for labeling. The result is promising --- 30 to 50\% of manual labeling cost can be saved with negligible impact on evaluation accuracy. 
However, reducing the amount of labels does not help improve label quality.
Despite less effort in this direction so far, we have noticed that there is an increasing attention to the general problem of data labeling in the SE community. A case study by Amershi et al.~\cite{amershi2019software} recognized data collection, cleaning, and labeling as crucial steps in the development pipeline of machine learning applications.
Other communities such as database and computer vision have proposed several supervised or semi-supervised learning techniques to automatically label data~\cite{benato2021semi, ke2019end, chen2014beat, wu2018tagging, ke2017data}. 
We suggest that readers refer to Roh et al.~\cite{roh2019survey} for a detailed survey of data collection and labeling tools built by other research communities. 
In the future, we believe it is worthwhile developing more automated or semi-automated tools such as \cite{kim2020reducing} to support reducing data labeling efforts in ADSs.




Furthermore, SE researchers can also contribute to data cleaning by developing input validation or outlier detection techniques on driving data. Traditional data cleaning approaches need users to provide a set of specific rules to determine constraints~\cite{gudivada2017data, chu2016data, rahm2000data}. Such rules are easy to define on tabular data or text data. However, it is challenging to define such rules for sensor data such as videos and point clouds that involve complex driving scenes. The SE community has proposed many techniques to identify failure-inducing inputs in software testing and debugging~\cite{zeller2002simplifying, misherghi2006hdd, clause2009penumbra, gulzar2016bigdebug, sun2018perses, gopinath2020abstracting}. For example, Gulzar et al.~\cite{gulzar2016bigdebug} leveraged data provenance to trace error propagation in big data systems and identify fault-inducing data. It would be beneficial to investigate how to adapt or renovate these techniques to identify low-quality data that leads to abnormal driving behavior in ADSs.





\section{Threats to validity} \label{sec:validity}
We discuss the threats to validity following the standard proposed by Wohlin et al.~\cite{Experimentation}.

\textbf{External Validity.} 
The external validity concerns about the generalizability of our findings. To mitigate the threat to external validity, we improved the diversity of the interviewees as much as possible. As shown in Section 2.1, these 10 interviewees came from 10 different companies and had different responsibilities. Furthermore, to expand the scale of our research, we also conducted 100 surveys, and these respondents were in a variety of positions, and came from various types of companies and organizations, as shown in Section 2.2. In addition, ADS developed and tested by these interviewees and questionnaire respondents were from \change{level 1 to level 5 \cite{sae2014taxonomy}}, which covered all levels of industrial ADS. Therefore, the interviewees and survey respondents in this study can represent the industrial participants of ADS testing.

\textbf{Construct Validity.} 
The construct validity concerns about the design of our 
\change{empirical study and literature review}. To mitigate this threat, for interviews, we first conducted two pilot interviews. Based on feedback from these two pilot interviews, we refined the interview guide and interview questions. In addition, to not bias participants' responses during the interview, we conducted semi-structured interviews and asked open-ended questions, which allowed for new ideas and questions to be brought up during the discussion. Though the survey questions were mainly multiple-choice questions, we allowed respondents to provide alternative answers. Also open-ended questions were added in the survey for respondents to supplement questions and answers not included in the survey. \change{In the literature review, we used broad keywords to include as many relevant papers as possible. Additionally, with manual filtering, we removed papers that were not relevant to autonomous driving testing. Though in the literature review process, we did not apply literature quality assessment as described in~\cite{keele2007guidelines}, the quality of searched papers can be assured because they are all published on high quality SE conferences and journals. We did not create detailed data collection forms for searched papers, which might affect the categorization of reviewed papers. The process of data collection will be refined in the future work.}
 
\textbf{Internal Validity.} 
The internal validity concerns about how accurate participants' responses may reflect their real needs and whether there are other compounding factors that may affect the results. Specifically, since interview participants were developers from technology companies, some of them may not faithfully reveal the exact and most emergent needs from their team due to confidentiality concerns. When designing the interview, we tried our best to avoid conversations that may involve technical details or make them feel uncomfortable to answer. In addition, before each interview, we shared the interview guide with the participant and informed them that our goal was to solicit high-level feedback rather than technical details. To avoid mistakes and biases in the thematic analysis step, the first and second authors carried out this step separately. Then two authors continuously adjusted the extraction result, until the inter-rater agreements between the two coders reached 0.85, in Cohen’s Kappa \cite{viera2005understanding}. Also, the collected information was regularly reported and discussed in the whole research team.

\section{conclusion} \label{sec:conclusion}
This paper presents an empirical study on the current practices and needs of testing autonomous driving systems. Through an interview study and a large-scale survey, we identified seven common practices and four common needs of ADS practitioners. 
These findings provide a deep understanding about how ADS practitioners test autonomous driving systems in practice and what kinds of tool support they find helpful. Furthermore, we did a literature review of related ADS testing research from the SE community and assessed to what extent existing work can address those industrial needs. We found that, though one need, {\em identifying possible corner cases and unexpected driving scenarios}, is widely investigated by the SE community, the other three needs are under-investigated. Furthermore, most existing work is designed for end-to-end driving models, while multi-module driving systems are widely adopted nowadays. Based on these gaps, we proposed four future directions to build better tool support for testing autonomous driving systems. We believe there are many research opportunities for SE researchers and therefore call for more attention and effort towards these future directions.

\section{Acknowledgements}
This work is in part supported by an Australian Research Council (ARC) Discovery Project (DP210102447), an ARC Linkage Project (LP190100676), and a DATA61 project (Data61 CRP C020996).

\bibliographystyle{ACM-Reference-Format}
\bibliography{reference.bib}


\begin{thebibliography}{111}


\ifx \showCODEN    \undefined \def \showCODEN     #1{\unskip}     \fi
\ifx \showDOI      \undefined \def \showDOI       #1{#1}\fi
\ifx \showISBNx    \undefined \def \showISBNx     #1{\unskip}     \fi
\ifx \showISBNxiii \undefined \def \showISBNxiii  #1{\unskip}     \fi
\ifx \showISSN     \undefined \def \showISSN      #1{\unskip}     \fi
\ifx \showLCCN     \undefined \def \showLCCN      #1{\unskip}     \fi
\ifx \shownote     \undefined \def \shownote      #1{#1}          \fi
\ifx \showarticletitle \undefined \def \showarticletitle #1{#1}   \fi
\ifx \showURL      \undefined \def \showURL       {\relax}        \fi
\providecommand\bibfield[2]{#2}
\providecommand\bibinfo[2]{#2}
\providecommand\natexlab[1]{#1}
\providecommand\showeprint[2][]{arXiv:#2}

\bibitem[Abdessalem et~al\mbox{.}(2018a)]%
        {abdessalem2018vision}
\bibfield{author}{\bibinfo{person}{Raja~Ben Abdessalem}, \bibinfo{person}{Shiva
  Nejati}, \bibinfo{person}{Lionel~C. Briand}, {and} \bibinfo{person}{Thomas
  Stifter}.} \bibinfo{year}{2018}\natexlab{a}.
\newblock \showarticletitle{Testing vision-based control systems using
  learnable evolutionary algorithms}. In \bibinfo{booktitle}{\emph{Proceedings
  of the 40th International Conference on Software Engineering, {ICSE} 2018,
  Gothenburg, Sweden, May 27 - June 03, 2018}},
  \bibfield{editor}{\bibinfo{person}{Michel Chaudron}, \bibinfo{person}{Ivica
  Crnkovic}, \bibinfo{person}{Marsha Chechik}, {and} \bibinfo{person}{Mark
  Harman}} (Eds.). \bibinfo{publisher}{{ACM}}, \bibinfo{pages}{1016--1026}.
\newblock
\urldef\tempurl%
\url{https://doi.org/10.1145/3180155.3180160}
\showDOI{\tempurl}


\bibitem[Abdessalem et~al\mbox{.}(2018b)]%
        {abdessalem2018testing}
\bibfield{author}{\bibinfo{person}{Raja~Ben Abdessalem},
  \bibinfo{person}{Annibale Panichella}, \bibinfo{person}{Shiva Nejati},
  \bibinfo{person}{Lionel~C Briand}, {and} \bibinfo{person}{Thomas Stifter}.}
  \bibinfo{year}{2018}\natexlab{b}.
\newblock \showarticletitle{Testing autonomous cars for feature interaction
  failures using many-objective search}. In \bibinfo{booktitle}{\emph{2018 33rd
  IEEE/ACM International Conference on Automated Software Engineering (ASE)}}.
  IEEE, \bibinfo{pages}{143--154}.
\newblock


\bibitem[Althoff et~al\mbox{.}(2017)]%
        {althoff2017commonroad}
\bibfield{author}{\bibinfo{person}{M. Althoff}, \bibinfo{person}{M. Koschi},
  {and} \bibinfo{person}{S. Manzinger}.} \bibinfo{year}{2017}\natexlab{}.
\newblock \showarticletitle{CommonRoad: Composable benchmarks for motion
  planning on roads}. In \bibinfo{booktitle}{\emph{IV}}. IEEE,
  \bibinfo{pages}{719--726}.
\newblock


\bibitem[Amershi et~al\mbox{.}(2019)]%
        {amershi2019software}
\bibfield{author}{\bibinfo{person}{S. Amershi}, \bibinfo{person}{A. Begel},
  \bibinfo{person}{C. Bird}, \bibinfo{person}{R. DeLine}, \bibinfo{person}{H.
  Gall}, \bibinfo{person}{E. Kamar}, \bibinfo{person}{N. Nagappan},
  \bibinfo{person}{B. Nushi}, {and} \bibinfo{person}{T. Zimmermann}.}
  \bibinfo{year}{2019}\natexlab{}.
\newblock \showarticletitle{Software engineering for machine learning: A case
  study}. In \bibinfo{booktitle}{\emph{ICSE}}. IEEE, \bibinfo{pages}{291--300}.
\newblock


\bibitem[ApolloAuto(2021)]%
        {Apollo}
\bibfield{author}{\bibinfo{person}{ApolloAuto}.}
  \bibinfo{year}{2021}\natexlab{}.
\newblock \bibinfo{title}{Apollo}.
\newblock \bibinfo{howpublished}{\url{https://bit.ly/2E3vWyo}}.
\newblock


\bibitem[Arcaini et~al\mbox{.}(2021)]%
        {arcaini2021targeting}
\bibfield{author}{\bibinfo{person}{Paolo Arcaini}, \bibinfo{person}{Xiao-Yi
  Zhang}, {and} \bibinfo{person}{Fuyuki Ishikawa}.}
  \bibinfo{year}{2021}\natexlab{}.
\newblock \showarticletitle{Targeting patterns of driving characteristics in
  testing autonomous driving systems}. In \bibinfo{booktitle}{\emph{ICST}}.
  IEEE, \bibinfo{pages}{295--305}.
\newblock


\bibitem[ASAM(2021)]%
        {OpenScenario}
\bibfield{author}{\bibinfo{person}{ASAM}.} \bibinfo{year}{2021}\natexlab{}.
\newblock \bibinfo{title}{ASAM OpenSCENARIO}.
\newblock \bibinfo{howpublished}{\url{https://bit.ly/3ya34Rm}}.
\newblock


\bibitem[Autoware-AI(2020)]%
        {autoware}
\bibfield{author}{\bibinfo{person}{Autoware-AI}.}
  \bibinfo{year}{2020}\natexlab{}.
\newblock \bibinfo{title}{autoware.ai}.
\newblock \bibinfo{howpublished}{\url{https://bit.ly/3gZ1gBS}}.
\newblock


\bibitem[Benato(2021)]%
        {benato2021semi}
\bibfield{author}{\bibinfo{person}{B. Benato}.}
  \bibinfo{year}{2021}\natexlab{}.
\newblock \showarticletitle{Semi-automatic data annotation guided by feature
  space projection}.
\newblock \bibinfo{journal}{\emph{Pattern Recognition}}  \bibinfo{volume}{109}
  (\bibinfo{year}{2021}), \bibinfo{pages}{107612}.
\newblock


\bibitem[Bender et~al\mbox{.}(2014)]%
        {bender2014lanelets}
\bibfield{author}{\bibinfo{person}{P. Bender}, \bibinfo{person}{J. Ziegler},
  {and} \bibinfo{person}{C. Stiller}.} \bibinfo{year}{2014}\natexlab{}.
\newblock \showarticletitle{Lanelets: Efficient map representation for
  autonomous driving}. In \bibinfo{booktitle}{\emph{IV}}. IEEE,
  \bibinfo{pages}{420--425}.
\newblock


\bibitem[Benekohal and Treiterer(1988)]%
        {benekohal1988carsim}
\bibfield{author}{\bibinfo{person}{R. Benekohal} {and} \bibinfo{person}{J.
  Treiterer}.} \bibinfo{year}{1988}\natexlab{}.
\newblock \showarticletitle{CARSIM: Car-following model for simulation of
  traffic in normal and stop-and-go conditions}.
\newblock \bibinfo{journal}{\emph{Transportation research record}}
  \bibinfo{volume}{1194} (\bibinfo{year}{1988}), \bibinfo{pages}{99--111}.
\newblock


\bibitem[Board(2019)]%
        {ADSAccident}
\bibfield{author}{\bibinfo{person}{The National Transportation~Safety Board}.}
  \bibinfo{year}{2019}\natexlab{}.
\newblock \bibinfo{title}{Preliminary Report Highway Hwy18mh010}.
\newblock \bibinfo{howpublished}{\url{https://bit.ly/2N0SHuj}}.
\newblock


\bibitem[Bojarski et~al\mbox{.}(2017)]%
        {bojarski2017explaining}
\bibfield{author}{\bibinfo{person}{M. Bojarski}, \bibinfo{person}{P. Yeres},
  \bibinfo{person}{A. Choromanska}, \bibinfo{person}{K. Choromanski},
  \bibinfo{person}{B. Firner}, \bibinfo{person}{Lawrence~D. Jackel}, {and}
  \bibinfo{person}{U. Muller}.} \bibinfo{year}{2017}\natexlab{}.
\newblock \showarticletitle{Explaining How a Deep Neural Network Trained with
  End-to-End Learning Steers a Car}.
\newblock \bibinfo{journal}{\emph{CoRR}}  \bibinfo{volume}{abs/1704.07911}
  (\bibinfo{year}{2017}).
\newblock


\bibitem[Borg et~al\mbox{.}(2021)]%
        {borg2021digital}
\bibfield{author}{\bibinfo{person}{Markus Borg}, \bibinfo{person}{Raja~Ben
  Abdessalem}, \bibinfo{person}{Shiva Nejati},
  \bibinfo{person}{Fran{\c{c}}ois-Xavier Jegeden}, {and}
  \bibinfo{person}{Donghwan Shin}.} \bibinfo{year}{2021}\natexlab{}.
\newblock \showarticletitle{Digital twins are not
  monozygotic--cross-replicating adas testing in two industry-grade automotive
  simulators}. In \bibinfo{booktitle}{\emph{ICST}}. IEEE,
  \bibinfo{pages}{383--393}.
\newblock


\bibitem[Borg et~al\mbox{.}(2018)]%
        {borg2018safely}
\bibfield{author}{\bibinfo{person}{M. Borg}, \bibinfo{person}{C. Englund},
  \bibinfo{person}{K. Wnuk}, \bibinfo{person}{B. Duran}, \bibinfo{person}{C.
  Levandowski}, \bibinfo{person}{S. Gao}, \bibinfo{person}{Y. Tan},
  \bibinfo{person}{H. Kaijser}, \bibinfo{person}{H. L{\"o}nn}, {and}
  \bibinfo{person}{J. T{\"o}rnqvist}.} \bibinfo{year}{2018}\natexlab{}.
\newblock \showarticletitle{Safely entering the deep: A review of verification
  and validation for machine learning and a challenge elicitation in the
  automotive industry}.
\newblock \bibinfo{journal}{\emph{arXiv preprint arXiv:1812.05389}}
  (\bibinfo{year}{2018}).
\newblock


\bibitem[Braun and Clarke(2006)]%
        {braun2006using}
\bibfield{author}{\bibinfo{person}{V. Braun} {and} \bibinfo{person}{V.
  Clarke}.} \bibinfo{year}{2006}\natexlab{}.
\newblock \showarticletitle{Using thematic analysis in psychology}.
\newblock \bibinfo{journal}{\emph{Qualitative research in psychology}}
  \bibinfo{volume}{3}, \bibinfo{number}{2} (\bibinfo{year}{2006}),
  \bibinfo{pages}{77--101}.
\newblock


\bibitem[Butler and Finelli(1991)]%
        {butler1991infeasibility}
\bibfield{author}{\bibinfo{person}{R. Butler} {and} \bibinfo{person}{G.
  Finelli}.} \bibinfo{year}{1991}\natexlab{}.
\newblock \showarticletitle{The infeasibility of experimental quantification of
  life-critical software reliability}. In \bibinfo{booktitle}{\emph{Conference
  on Software for Citical Systems}}. \bibinfo{pages}{66--76}.
\newblock


\bibitem[Cal{\`o} et~al\mbox{.}(2020)]%
        {calo2020generating}
\bibfield{author}{\bibinfo{person}{Alessandro Cal{\`o}}, \bibinfo{person}{Paolo
  Arcaini}, \bibinfo{person}{Shaukat Ali}, \bibinfo{person}{Florian Hauer},
  {and} \bibinfo{person}{Fuyuki Ishikawa}.} \bibinfo{year}{2020}\natexlab{}.
\newblock \showarticletitle{Generating avoidable collision scenarios for
  testing autonomous driving systems}. In \bibinfo{booktitle}{\emph{ICST}}.
  IEEE, \bibinfo{pages}{375--386}.
\newblock


\bibitem[Carla(2021)]%
        {carla}
\bibfield{author}{\bibinfo{person}{Carla}.} \bibinfo{year}{2021}\natexlab{}.
\newblock \bibinfo{title}{Carla: Open-Source Simulator for Autonomous Driving
  Research}.
\newblock \bibinfo{howpublished}{\url{https://bit.ly/3qE26qA}}.
\newblock


\bibitem[Chandrasekaran et~al\mbox{.}(2021)]%
        {chandrasekaran2021combinatorial}
\bibfield{author}{\bibinfo{person}{Jaganmohan Chandrasekaran},
  \bibinfo{person}{Yu Lei}, \bibinfo{person}{Raghu Kacker}, {and}
  \bibinfo{person}{D~Richard Kuhn}.} \bibinfo{year}{2021}\natexlab{}.
\newblock \showarticletitle{A Combinatorial Approach to Testing Deep Neural
  Network-based Autonomous Driving Systems}. In \bibinfo{booktitle}{\emph{2021
  IEEE International Conference on Software Testing, Verification and
  Validation Workshops (ICSTW)}}. IEEE, \bibinfo{pages}{57--66}.
\newblock


\bibitem[Chen et~al\mbox{.}(2014)]%
        {chen2014beat}
\bibfield{author}{\bibinfo{person}{L. Chen}, \bibinfo{person}{S. Fidler},
  \bibinfo{person}{A. Yuille}, {and} \bibinfo{person}{R. Urtasun}.}
  \bibinfo{year}{2014}\natexlab{}.
\newblock \showarticletitle{Beat the mturkers: Automatic image labeling from
  weak 3d supervision}. In \bibinfo{booktitle}{\emph{CVPR}}.
  \bibinfo{pages}{3198--3205}.
\newblock


\bibitem[Chu et~al\mbox{.}(2016)]%
        {chu2016data}
\bibfield{author}{\bibinfo{person}{X. Chu}, \bibinfo{person}{I. Ilyas},
  \bibinfo{person}{S. Krishnan}, {and} \bibinfo{person}{J. Wang}.}
  \bibinfo{year}{2016}\natexlab{}.
\newblock \showarticletitle{Data cleaning: Overview and emerging challenges}.
  In \bibinfo{booktitle}{\emph{International Conference on Management of
  Data}}. \bibinfo{pages}{2201--2206}.
\newblock


\bibitem[Clause et~al\mbox{.}(2009)]%
        {clause2009penumbra}
\bibfield{author}{\bibinfo{person}{J. Clause}, \bibinfo{person}{A. Orso}, {and}
  \bibinfo{person}{ro}.} \bibinfo{year}{2009}\natexlab{}.
\newblock \showarticletitle{Penumbra: Automatically identifying
  failure-relevant inputs using dynamic tainting}. In
  \bibinfo{booktitle}{\emph{ISSTA}}. \bibinfo{pages}{249--260}.
\newblock


\bibitem[comma. ai(2021)]%
        {openpilot}
\bibfield{author}{\bibinfo{person}{comma. ai}.}
  \bibinfo{year}{2021}\natexlab{}.
\newblock \bibinfo{title}{OpenPilot}.
\newblock \bibinfo{howpublished}{\url{https://bit.ly/3w099fI}}.
\newblock


\bibitem[Committee et~al\mbox{.}(2014)]%
        {sae2014taxonomy}
\bibfield{author}{\bibinfo{person}{SAE On-Road Automated Vehicle~Standards
  Committee} {et~al\mbox{.}}} \bibinfo{year}{2014}\natexlab{}.
\newblock \showarticletitle{Taxonomy and definitions for terms related to
  on-road motor vehicle automated driving systems}.
\newblock \bibinfo{journal}{\emph{SAE Standard J}}  \bibinfo{volume}{3016}
  (\bibinfo{year}{2014}), \bibinfo{pages}{1--16}.
\newblock


\bibitem[Cordts et~al\mbox{.}(2016)]%
        {Cordts2016Cityscapes}
\bibfield{author}{\bibinfo{person}{M. Cordts}, \bibinfo{person}{M. Omran},
  \bibinfo{person}{S. Ramos}, \bibinfo{person}{T. Rehfeld}, \bibinfo{person}{M.
  Enzweiler}, \bibinfo{person}{R. Benenson}, \bibinfo{person}{U. Franke},
  \bibinfo{person}{S. Roth}, {and} \bibinfo{person}{B. Schiele}.}
  \bibinfo{year}{2016}\natexlab{}.
\newblock \showarticletitle{The Cityscapes Dataset for Semantic Urban Scene
  Understanding}. In \bibinfo{booktitle}{\emph{CVPR}}.
\newblock


\bibitem[DeepDrive(2020)]%
        {DeepDrive}
\bibfield{author}{\bibinfo{person}{DeepDrive}.}
  \bibinfo{year}{2020}\natexlab{}.
\newblock \bibinfo{title}{DeepDrive}.
\newblock \bibinfo{howpublished}{\url{https://bit.ly/2OTsheJ}}.
\newblock


\bibitem[Deng et~al\mbox{.}(2020)]%
        {deng2020analysis}
\bibfield{author}{\bibinfo{person}{Yao Deng}, \bibinfo{person}{Xi Zheng},
  \bibinfo{person}{Tianyi Zhang}, \bibinfo{person}{Chen Chen},
  \bibinfo{person}{Guannan Lou}, {and} \bibinfo{person}{Miryung Kim}.}
  \bibinfo{year}{2020}\natexlab{}.
\newblock \showarticletitle{An analysis of adversarial attacks and defenses on
  autonomous driving models}. In \bibinfo{booktitle}{\emph{2020 IEEE
  international conference on pervasive computing and communications
  (PerCom)}}. IEEE, \bibinfo{pages}{1--10}.
\newblock


\bibitem[Dey and Lee(2021)]%
        {Dey_2021multilayered}
\bibfield{author}{\bibinfo{person}{S. Dey} {and} \bibinfo{person}{S. Lee}.}
  \bibinfo{year}{2021}\natexlab{}.
\newblock \showarticletitle{Multilayered review of safety approaches for
  machine learning-based systems in the days of {AI}}.
\newblock \bibinfo{journal}{\emph{Journal of Systems and Software}}
  \bibinfo{volume}{176} (\bibinfo{date}{jun} \bibinfo{year}{2021}),
  \bibinfo{pages}{110941}.
\newblock


\bibitem[Du et~al\mbox{.}(2021)]%
        {du2021towards}
\bibfield{author}{\bibinfo{person}{Dehui Du}, \bibinfo{person}{Jiena Chen},
  \bibinfo{person}{Mingzhuo Zhang}, {and} \bibinfo{person}{Mingjun Ma}.}
  \bibinfo{year}{2021}\natexlab{}.
\newblock \showarticletitle{Towards Verified Safety-critical Autonomous Driving
  Scenario with ADSML}. In \bibinfo{booktitle}{\emph{2021 IEEE 45th Annual
  Computers, Software, and Applications Conference (COMPSAC)}}. IEEE,
  \bibinfo{pages}{1333--1338}.
\newblock


\bibitem[Favar et~al\mbox{.}(2018)]%
        {favaro2018autonomous}
\bibfield{author}{\bibinfo{person}{F. Favar}, \bibinfo{person}{S. Eurich},
  {and} \bibinfo{person}{N. Nader}.} \bibinfo{year}{2018}\natexlab{}.
\newblock \showarticletitle{Autonomous vehicles’ disengagements: Trends,
  triggers, and regulatory limitations}.
\newblock \bibinfo{journal}{\emph{Accident Analysis \& Prevention}}
  \bibinfo{volume}{110} (\bibinfo{year}{2018}), \bibinfo{pages}{136--148}.
\newblock


\bibitem[Feng et~al\mbox{.}(2020)]%
        {feng2020deepGini}
\bibfield{author}{\bibinfo{person}{Y. Feng}, \bibinfo{person}{Q. Shi},
  \bibinfo{person}{X. Gao}, \bibinfo{person}{J. Wan}, \bibinfo{person}{C.
  Fang}, {and} \bibinfo{person}{Z. Chen}.} \bibinfo{year}{2020}\natexlab{}.
\newblock \showarticletitle{DeepGini: prioritizing massive tests to enhance the
  robustness of deep neural networks}. In \bibinfo{booktitle}{\emph{ISSTA}}.
  \bibinfo{publisher}{{ACM}}, \bibinfo{pages}{177--188}.
\newblock


\bibitem[Fremont et~al\mbox{.}(2019)]%
        {fremont2019scenic}
\bibfield{author}{\bibinfo{person}{D. Fremont}, \bibinfo{person}{T. Dreossi},
  \bibinfo{person}{S. Ghosh}, \bibinfo{person}{X. Yue}, \bibinfo{person}{A.
  Sangiovanni-Vincentelli}, {and} \bibinfo{person}{S. Seshia}.}
  \bibinfo{year}{2019}\natexlab{}.
\newblock \showarticletitle{Scenic: a language for scenario specification and
  scene generation}. In \bibinfo{booktitle}{\emph{PLDI}}.
  \bibinfo{pages}{63--78}.
\newblock


\bibitem[Fritzsch et~al\mbox{.}(2021)]%
        {fritzsch2021experiences}
\bibfield{author}{\bibinfo{person}{Jonas Fritzsch}, \bibinfo{person}{Tobias
  Schmid}, {and} \bibinfo{person}{Stefan Wagner}.}
  \bibinfo{year}{2021}\natexlab{}.
\newblock \showarticletitle{Experiences from Large-Scale Model Checking:
  Verifying a Vehicle Control System with NuSMV}. In
  \bibinfo{booktitle}{\emph{ICST}}. IEEE, \bibinfo{pages}{372--382}.
\newblock


\bibitem[Gambi et~al\mbox{.}(2019a)]%
        {gambi2019reconstructing}
\bibfield{author}{\bibinfo{person}{Alessio Gambi}, \bibinfo{person}{Tri Huynh},
  {and} \bibinfo{person}{Gordon Fraser}.} \bibinfo{year}{2019}\natexlab{a}.
\newblock \showarticletitle{Automatically reconstructing car crashes from
  police reports for testing self-driving cars}. In
  \bibinfo{booktitle}{\emph{2019 IEEE/ACM 41st International Conference on
  Software Engineering: Companion Proceedings (ICSE-Companion)}}. IEEE,
  \bibinfo{pages}{290--291}.
\newblock


\bibitem[Gambi et~al\mbox{.}(2019b)]%
        {gambi2019generating}
\bibfield{author}{\bibinfo{person}{Alessio Gambi}, \bibinfo{person}{Tri Huynh},
  {and} \bibinfo{person}{Gordon Fraser}.} \bibinfo{year}{2019}\natexlab{b}.
\newblock \showarticletitle{Generating effective test cases for self-driving
  cars from police reports}. In \bibinfo{booktitle}{\emph{Proceedings of the
  2019 27th ACM Joint Meeting on European Software Engineering Conference and
  Symposium on the Foundations of Software Engineering}}.
  \bibinfo{pages}{257--267}.
\newblock


\bibitem[Gambi et~al\mbox{.}(2019c)]%
        {gambi2019automatically}
\bibfield{author}{\bibinfo{person}{Alessio Gambi}, \bibinfo{person}{Marc
  Mueller}, {and} \bibinfo{person}{Gordon Fraser}.}
  \bibinfo{year}{2019}\natexlab{c}.
\newblock \showarticletitle{Automatically testing self-driving cars with
  search-based procedural content generation}. In
  \bibinfo{booktitle}{\emph{Proceedings of the 28th ACM SIGSOFT International
  Symposium on Software Testing and Analysis}}. \bibinfo{pages}{318--328}.
\newblock


\bibitem[Gambi et~al\mbox{.}(2019d)]%
        {gambi2019asfault}
\bibfield{author}{\bibinfo{person}{Alessio Gambi}, \bibinfo{person}{Marc
  M{\"u}ller}, {and} \bibinfo{person}{Gordon Fraser}.}
  \bibinfo{year}{2019}\natexlab{d}.
\newblock \showarticletitle{Asfault: Testing self-driving car software using
  search-based procedural content generation}. In
  \bibinfo{booktitle}{\emph{2019 IEEE/ACM 41st International Conference on
  Software Engineering: Companion Proceedings (ICSE-Companion)}}. IEEE,
  \bibinfo{pages}{27--30}.
\newblock


\bibitem[Garcia et~al\mbox{.}(2020)]%
        {garcia2020comprehensive}
\bibfield{author}{\bibinfo{person}{J. Garcia}, \bibinfo{person}{Y. Feng},
  \bibinfo{person}{J. Shen}, \bibinfo{person}{S. Almanee}, \bibinfo{person}{Y.
  Xia}, \bibinfo{person}{Chen}, {and} \bibinfo{person}{Q. Alfred}.}
  \bibinfo{year}{2020}\natexlab{}.
\newblock \showarticletitle{A comprehensive study of autonomous vehicle bugs}.
  In \bibinfo{booktitle}{\emph{ICSE}}. \bibinfo{pages}{385--396}.
\newblock


\bibitem[Gibbs(2017)]%
        {gibbs2017google}
\bibfield{author}{\bibinfo{person}{S. Gibbs}.} \bibinfo{year}{2017}\natexlab{}.
\newblock \showarticletitle{Google sibling waymo launches fully autonomous
  ride-hailing service}.
\newblock \bibinfo{journal}{\emph{The Guardian}}  \bibinfo{volume}{7}
  (\bibinfo{year}{2017}).
\newblock


\bibitem[Gladisch et~al\mbox{.}(2019)]%
        {gladisch2019experience}
\bibfield{author}{\bibinfo{person}{Christoph Gladisch}, \bibinfo{person}{Thomas
  Heinz}, \bibinfo{person}{Christian Heinzemann}, \bibinfo{person}{Jens
  Oehlerking}, \bibinfo{person}{Anne von Vietinghoff}, {and}
  \bibinfo{person}{Tim Pfitzer}.} \bibinfo{year}{2019}\natexlab{}.
\newblock \showarticletitle{Experience paper: Search-based testing in automated
  driving control applications}. In \bibinfo{booktitle}{\emph{2019 34th
  IEEE/ACM International Conference on Automated Software Engineering (ASE)}}.
  IEEE, \bibinfo{pages}{26--37}.
\newblock


\bibitem[Gopinath et~al\mbox{.}(2020)]%
        {gopinath2020abstracting}
\bibfield{author}{\bibinfo{person}{R. Gopinath}, \bibinfo{person}{A. Kampmann},
  \bibinfo{person}{er}, \bibinfo{person}{N. Havrikov}, \bibinfo{person}{E.
  Soremekun}, {and} \bibinfo{person}{A. Zeller}.}
  \bibinfo{year}{2020}\natexlab{}.
\newblock \showarticletitle{Abstracting failure-inducing inputs}. In
  \bibinfo{booktitle}{\emph{ISSTA}}. \bibinfo{pages}{237--248}.
\newblock


\bibitem[Gudivada et~al\mbox{.}(2017)]%
        {gudivada2017data}
\bibfield{author}{\bibinfo{person}{V. Gudivada}, \bibinfo{person}{A. Apon},
  {and} \bibinfo{person}{J. Ding}.} \bibinfo{year}{2017}\natexlab{}.
\newblock \showarticletitle{Data quality considerations for big data and
  machine learning: Going beyond data cleaning and transformations}.
\newblock \bibinfo{journal}{\emph{International Journal on Advances in
  Software}} \bibinfo{volume}{10}, \bibinfo{number}{1} (\bibinfo{year}{2017}),
  \bibinfo{pages}{1--20}.
\newblock


\bibitem[Gulzar et~al\mbox{.}(2016)]%
        {gulzar2016bigdebug}
\bibfield{author}{\bibinfo{person}{M. Gulzar}, \bibinfo{person}{Interl},
  \bibinfo{person}{M. i}, \bibinfo{person}{S. Yoo}, \bibinfo{person}{S.
  Tetali}, \bibinfo{person}{T. Condie}, \bibinfo{person}{T. Millstein}, {and}
  \bibinfo{person}{M. Kim}.} \bibinfo{year}{2016}\natexlab{}.
\newblock \showarticletitle{Bigdebug: Debugging primitives for interactive big
  data processing in spark}. In \bibinfo{booktitle}{\emph{ICSE}}. IEEE,
  \bibinfo{pages}{784--795}.
\newblock


\bibitem[Han and Zhou(2020)]%
        {han2020metamorphic}
\bibfield{author}{\bibinfo{person}{J. Han} {and} \bibinfo{person}{Z. Zhou}.}
  \bibinfo{year}{2020}\natexlab{}.
\newblock \showarticletitle{Metamorphic Fuzz Testing of Autonomous Vehicles}.
  In \bibinfo{booktitle}{\emph{ICSE Workshop}}. \bibinfo{publisher}{{ACM}},
  \bibinfo{pages}{380--385}.
\newblock


\bibitem[Haq et~al\mbox{.}(2020)]%
        {haq2020comparing}
\bibfield{author}{\bibinfo{person}{Fitash~Ul Haq}, \bibinfo{person}{Donghwan
  Shin}, \bibinfo{person}{Shiva Nejati}, {and} \bibinfo{person}{Lionel~C
  Briand}.} \bibinfo{year}{2020}\natexlab{}.
\newblock \showarticletitle{Comparing offline and online testing of deep neural
  networks: An autonomous car case study}. In \bibinfo{booktitle}{\emph{ICST}}.
  IEEE, \bibinfo{pages}{85--95}.
\newblock


\bibitem[Hasler(2022)]%
        {tesla}
\bibfield{author}{\bibinfo{person}{Arthur~Frederick Hasler}.}
  \bibinfo{year}{2022}\natexlab{}.
\newblock \bibinfo{title}{60,000 Drivers Now Have Tesla Full Self Driving
  (FSD)---What It Is \& How To Get It}.
\newblock \bibinfo{howpublished}{\url{https://bit.ly/3t1zIST}}.
\newblock


\bibitem[Hu et~al\mbox{.}(2021)]%
        {hu2021disclosing}
\bibfield{author}{\bibinfo{person}{Zhisheng Hu}, \bibinfo{person}{Shengjian
  Guo}, \bibinfo{person}{Zhenyu Zhong}, {and} \bibinfo{person}{Kang Li}.}
  \bibinfo{year}{2021}\natexlab{}.
\newblock \showarticletitle{Disclosing the Fragility Problem of Virtual Safety
  Testing for Autonomous Driving Systems}. In \bibinfo{booktitle}{\emph{2021
  IEEE International Symposium on Software Reliability Engineering Workshops
  (ISSREW)}}. IEEE, \bibinfo{pages}{387--392}.
\newblock


\bibitem[Huang et~al\mbox{.}(2016)]%
        {huang2016autonomous}
\bibfield{author}{\bibinfo{person}{W. Huang}, \bibinfo{person}{K. Wang},
  \bibinfo{person}{Y. Lv}, {and} \bibinfo{person}{F. Zhu}.}
  \bibinfo{year}{2016}\natexlab{}.
\newblock \showarticletitle{Autonomous vehicles testing methods review}. In
  \bibinfo{booktitle}{\emph{ITSC}}. IEEE, \bibinfo{pages}{163--168}.
\newblock


\bibitem[Ishikawa(2020)]%
        {ishikawa2020testing}
\bibfield{author}{\bibinfo{person}{Fuyuki Ishikawa}.}
  \bibinfo{year}{2020}\natexlab{}.
\newblock \showarticletitle{Testing and Debugging Autonomous Driving:
  Experiences with Path Planner and Future Challenges}. In
  \bibinfo{booktitle}{\emph{2020 IEEE International Symposium on Software
  Reliability Engineering Workshops (ISSREW)}}. IEEE,
  \bibinfo{pages}{xxxiii--xxxiv}.
\newblock


\bibitem[Jahangirova et~al\mbox{.}(2021)]%
        {jahangirova2021quality}
\bibfield{author}{\bibinfo{person}{Gunel Jahangirova}, \bibinfo{person}{Andrea
  Stocco}, {and} \bibinfo{person}{Paolo Tonella}.}
  \bibinfo{year}{2021}\natexlab{}.
\newblock \showarticletitle{Quality metrics and oracles for autonomous vehicles
  testing}. In \bibinfo{booktitle}{\emph{ICST}}. IEEE,
  \bibinfo{pages}{194--204}.
\newblock


\bibitem[Johnson and Onwuegbuzie(2004)]%
        {johnson2004mixed}
\bibfield{author}{\bibinfo{person}{R~Burke Johnson} {and}
  \bibinfo{person}{Anthony~J Onwuegbuzie}.} \bibinfo{year}{2004}\natexlab{}.
\newblock \showarticletitle{Mixed methods research: A research paradigm whose
  time has come}.
\newblock \bibinfo{journal}{\emph{Educational researcher}}
  \bibinfo{volume}{33}, \bibinfo{number}{7} (\bibinfo{year}{2004}),
  \bibinfo{pages}{14--26}.
\newblock


\bibitem[Joshi(2020)]%
        {joshi2020amazon}
\bibfield{author}{\bibinfo{person}{A. Joshi}.} \bibinfo{year}{2020}\natexlab{}.
\newblock \showarticletitle{Amazon’s machine learning toolkit: Sagemaker}.
\newblock In \bibinfo{booktitle}{\emph{Machine Learning and Artificial
  Intelligence}}. \bibinfo{publisher}{Springer}, \bibinfo{pages}{233--243}.
\newblock


\bibitem[Kato et~al\mbox{.}(2018)]%
        {kato2018autoware}
\bibfield{author}{\bibinfo{person}{S. Kato}, \bibinfo{person}{S. Tokunaga},
  \bibinfo{person}{Y. Maruyama}, \bibinfo{person}{S. Maeda},
  \bibinfo{person}{M. Hirabayashi}, \bibinfo{person}{Y. Kitsukawa},
  \bibinfo{person}{A. Monrroy}, \bibinfo{person}{T. Ando}, \bibinfo{person}{Y.
  Fujii}, {and} \bibinfo{person}{T. Azumi}.} \bibinfo{year}{2018}\natexlab{}.
\newblock \showarticletitle{Autoware on board: Enabling autonomous vehicles
  with embedded systems}. In \bibinfo{booktitle}{\emph{ICCPS}}. IEEE,
  \bibinfo{pages}{287--296}.
\newblock


\bibitem[Ke et~al\mbox{.}(2017)]%
        {ke2017data}
\bibfield{author}{\bibinfo{person}{X. Ke}, \bibinfo{person}{M. Zhou},
  \bibinfo{person}{Y. Niu}, {and} \bibinfo{person}{W. Guo}.}
  \bibinfo{year}{2017}\natexlab{}.
\newblock \showarticletitle{Data equilibrium based automatic image annotation
  by fusing deep model and semantic propagation}.
\newblock \bibinfo{journal}{\emph{Pattern Recognition}}  \bibinfo{volume}{71}
  (\bibinfo{year}{2017}), \bibinfo{pages}{60--77}.
\newblock


\bibitem[Ke et~al\mbox{.}(2019)]%
        {ke2019end}
\bibfield{author}{\bibinfo{person}{X. Ke}, \bibinfo{person}{J. Zou}, {and}
  \bibinfo{person}{Y. Niu}.} \bibinfo{year}{2019}\natexlab{}.
\newblock \showarticletitle{End-to-end automatic image annotation based on deep
  CNN and multi-label data augmentation}.
\newblock \bibinfo{journal}{\emph{IEEE Transactions on Multimedia}}
  \bibinfo{volume}{21}, \bibinfo{number}{8} (\bibinfo{year}{2019}),
  \bibinfo{pages}{2093--2106}.
\newblock


\bibitem[Keele et~al\mbox{.}(2007)]%
        {keele2007guidelines}
\bibfield{author}{\bibinfo{person}{Staffs Keele} {et~al\mbox{.}}}
  \bibinfo{year}{2007}\natexlab{}.
\newblock \bibinfo{booktitle}{\emph{Guidelines for performing systematic
  literature reviews in software engineering}}.
\newblock \bibinfo{type}{{T}echnical {R}eport}. \bibinfo{institution}{Technical
  report, ver. 2.3 ebse technical report. ebse}.
\newblock


\bibitem[Kim et~al\mbox{.}(2020)]%
        {kim2020reducing}
\bibfield{author}{\bibinfo{person}{Jinhan Kim}, \bibinfo{person}{Jeongil Ju},
  \bibinfo{person}{Robert Feldt}, {and} \bibinfo{person}{Shin Yoo}.}
  \bibinfo{year}{2020}\natexlab{}.
\newblock \showarticletitle{Reducing dnn labelling cost using surprise
  adequacy: An industrial case study for autonomous driving}. In
  \bibinfo{booktitle}{\emph{Proceedings of the 28th ACM Joint Meeting on
  European Software Engineering Conference and Symposium on the Foundations of
  Software Engineering}}. \bibinfo{pages}{1466--1476}.
\newblock


\bibitem[King et~al\mbox{.}(2019)]%
        {king2019automated}
\bibfield{author}{\bibinfo{person}{Christian King}, \bibinfo{person}{Lennart
  Ries}, \bibinfo{person}{Christopher Kober}, \bibinfo{person}{Christoph
  Wohlfahrt}, {and} \bibinfo{person}{Eric Sax}.}
  \bibinfo{year}{2019}\natexlab{}.
\newblock \showarticletitle{Automated function assessment in driving
  scenarios}. In \bibinfo{booktitle}{\emph{ICST}}. IEEE,
  \bibinfo{pages}{414--419}.
\newblock


\bibitem[Kl{\"u}ck et~al\mbox{.}(2018)]%
        {kluck2018using}
\bibfield{author}{\bibinfo{person}{Florian Kl{\"u}ck}, \bibinfo{person}{Yihao
  Li}, \bibinfo{person}{Mihai Nica}, \bibinfo{person}{Jianbo Tao}, {and}
  \bibinfo{person}{Franz Wotawa}.} \bibinfo{year}{2018}\natexlab{}.
\newblock \showarticletitle{Using ontologies for test suites generation for
  automated and autonomous driving functions}. In
  \bibinfo{booktitle}{\emph{2018 IEEE International symposium on software
  reliability engineering workshops (ISSREW)}}. IEEE,
  \bibinfo{pages}{118--123}.
\newblock


\bibitem[Kluck et~al\mbox{.}(2019)]%
        {kluck2019genetic}
\bibfield{author}{\bibinfo{person}{F. Kluck}, \bibinfo{person}{M. Zimmermann},
  \bibinfo{person}{F. Wotawa}, {and} \bibinfo{person}{M. Nica}.}
  \bibinfo{year}{2019}\natexlab{}.
\newblock \showarticletitle{Genetic Algorithm-Based Test Parameter Optimization
  for {ADAS} System Testing}. In \bibinfo{booktitle}{\emph{QRS}}.
  \bibinfo{publisher}{{IEEE}}.
\newblock


\bibitem[Knauss et~al\mbox{.}(2017)]%
        {knauss2017software}
\bibfield{author}{\bibinfo{person}{Alessia Knauss}, \bibinfo{person}{Jan
  Schroder}, \bibinfo{person}{Christian Berger}, {and} \bibinfo{person}{Henrik
  Eriksson}.} \bibinfo{year}{2017}\natexlab{}.
\newblock \showarticletitle{Software-related challenges of testing automated
  vehicles}. In \bibinfo{booktitle}{\emph{2017 IEEE/ACM 39th International
  Conference on Software Engineering Companion (ICSE-C)}}. IEEE,
  \bibinfo{pages}{328--330}.
\newblock


\bibitem[Koopman and Wagner(2016)]%
        {koopman2016challenges}
\bibfield{author}{\bibinfo{person}{P. Koopman} {and} \bibinfo{person}{M.
  Wagner}.} \bibinfo{year}{2016}\natexlab{}.
\newblock \showarticletitle{Challenges in autonomous vehicle testing and
  validation}.
\newblock \bibinfo{journal}{\emph{SAE International Journal of Transportation
  Safety}} \bibinfo{volume}{4}, \bibinfo{number}{1} (\bibinfo{year}{2016}),
  \bibinfo{pages}{15--24}.
\newblock


\bibitem[Leudet et~al\mbox{.}(2019)]%
        {leudet2019ailivesim}
\bibfield{author}{\bibinfo{person}{J{\'e}r{\^o}me Leudet},
  \bibinfo{person}{Fran{\c{c}}ois Christophe}, \bibinfo{person}{Tommi
  Mikkonen}, {and} \bibinfo{person}{Tomi M{\"a}nnist{\"o}}.}
  \bibinfo{year}{2019}\natexlab{}.
\newblock \showarticletitle{Ailivesim: An extensible virtual environment for
  training autonomous vehicles}. In \bibinfo{booktitle}{\emph{2019 IEEE 43rd
  annual computer software and applications conference (COMPSAC)}},
  Vol.~\bibinfo{volume}{1}. IEEE, \bibinfo{pages}{479--488}.
\newblock


\bibitem[lgsvl(2020)]%
        {lgsvl}
\bibfield{author}{\bibinfo{person}{lgsvl}.} \bibinfo{year}{2020}\natexlab{}.
\newblock \bibinfo{title}{simulator}.
\newblock \bibinfo{howpublished}{\url{https://bit.ly/3dBDif1}}.
\newblock


\bibitem[Li et~al\mbox{.}(2020a)]%
        {li2020av}
\bibfield{author}{\bibinfo{person}{Guanpeng Li}, \bibinfo{person}{Yiran Li},
  \bibinfo{person}{Saurabh Jha}, \bibinfo{person}{Timothy Tsai},
  \bibinfo{person}{Michael Sullivan}, \bibinfo{person}{Siva Kumar~Sastry Hari},
  \bibinfo{person}{Zbigniew Kalbarczyk}, {and} \bibinfo{person}{Ravishankar
  Iyer}.} \bibinfo{year}{2020}\natexlab{a}.
\newblock \showarticletitle{AV-FUZZER: Finding safety violations in autonomous
  driving systems}. In \bibinfo{booktitle}{\emph{ISSRE}}. IEEE,
  \bibinfo{pages}{25--36}.
\newblock


\bibitem[Li et~al\mbox{.}(2020b)]%
        {li2020ontology}
\bibfield{author}{\bibinfo{person}{Yihao Li}, \bibinfo{person}{Jianbo Tao},
  {and} \bibinfo{person}{Franz Wotawa}.} \bibinfo{year}{2020}\natexlab{b}.
\newblock \showarticletitle{Ontology-based test generation for automated and
  autonomous driving functions}.
\newblock \bibinfo{journal}{\emph{Information and software technology}}
  \bibinfo{volume}{117} (\bibinfo{year}{2020}), \bibinfo{pages}{106200}.
\newblock


\bibitem[Li et~al\mbox{.}(2020c)]%
        {Li_2020PEACEPACT}
\bibfield{author}{\bibinfo{person}{Z. Li}, \bibinfo{person}{L. Zhang},
  \bibinfo{person}{J. Yan}, \bibinfo{person}{J. Zhang}, \bibinfo{person}{Z.
  Zhang}, {and} \bibinfo{person}{T.~H. Tse}.} \bibinfo{year}{2020}\natexlab{c}.
\newblock \showarticletitle{{PEACEPACT}: Prioritizing Examples to Accelerate
  Perturbation-Based Adversary Generation for {DNN} Classification Testing}. In
  \bibinfo{booktitle}{\emph{QRS}}. \bibinfo{publisher}{{IEEE}}.
\newblock


\bibitem[Liu and Capretz(2021)]%
        {liu2021analysis}
\bibfield{author}{\bibinfo{person}{Siyuan Liu} {and}
  \bibinfo{person}{Luiz~Fernando Capretz}.} \bibinfo{year}{2021}\natexlab{}.
\newblock \showarticletitle{An Analysis of Testing Scenarios for Automated
  Driving Systems}. In \bibinfo{booktitle}{\emph{2021 IEEE International
  Conference on Software Analysis, Evolution and Reengineering (SANER)}}. IEEE,
  \bibinfo{pages}{622--629}.
\newblock


\bibitem[Lu et~al\mbox{.}(2021)]%
        {lu2021search}
\bibfield{author}{\bibinfo{person}{Chengjie Lu}, \bibinfo{person}{Huihui
  Zhang}, \bibinfo{person}{Tao Yue}, {and} \bibinfo{person}{Shaukat Ali}.}
  \bibinfo{year}{2021}\natexlab{}.
\newblock \showarticletitle{Search-Based Selection and Prioritization of Test
  Scenarios for Autonomous Driving Systems}. In
  \bibinfo{booktitle}{\emph{International Symposium on Search Based Software
  Engineering}}. Springer, \bibinfo{pages}{41--55}.
\newblock


\bibitem[Luo et~al\mbox{.}(2021)]%
        {luo2021targeting}
\bibfield{author}{\bibinfo{person}{Yixing Luo}, \bibinfo{person}{Xiao-Yi
  Zhang}, \bibinfo{person}{Paolo Arcaini}, \bibinfo{person}{Zhi Jin},
  \bibinfo{person}{Haiyan Zhao}, \bibinfo{person}{Fuyuki Ishikawa},
  \bibinfo{person}{Rongxin Wu}, {and} \bibinfo{person}{Tao Xie}.}
  \bibinfo{year}{2021}\natexlab{}.
\newblock \showarticletitle{Targeting Requirements Violations of Autonomous
  Driving Systems by Dynamic Evolutionary Search}. In
  \bibinfo{booktitle}{\emph{2021 36th IEEE/ACM International Conference on
  Automated Software Engineering (ASE)}}. IEEE, \bibinfo{pages}{279--291}.
\newblock


\bibitem[Ma et~al\mbox{.}(2021)]%
        {ma2021test}
\bibfield{author}{\bibinfo{person}{W. Ma}, \bibinfo{person}{M. Papadakis},
  \bibinfo{person}{A. Tsakmalis}, \bibinfo{person}{M. Cordy}, {and}
  \bibinfo{person}{Y. Traon}.} \bibinfo{year}{2021}\natexlab{}.
\newblock \showarticletitle{Test selection for deep learning systems}.
\newblock \bibinfo{journal}{\emph{ACM Transactions on Software Engineering and
  Methodology (TOSEM)}} \bibinfo{volume}{30}, \bibinfo{number}{2}
  (\bibinfo{year}{2021}), \bibinfo{pages}{1--22}.
\newblock


\bibitem[Masuda(2017)]%
        {masuda2017software}
\bibfield{author}{\bibinfo{person}{Satoshi Masuda}.}
  \bibinfo{year}{2017}\natexlab{}.
\newblock \showarticletitle{Software testing design techniques used in
  automated vehicle simulations}. In \bibinfo{booktitle}{\emph{2017 IEEE
  International Conference on Software Testing, Verification and Validation
  Workshops (ICSTW)}}. IEEE, \bibinfo{pages}{300--303}.
\newblock


\bibitem[microsoft(2021a)]%
        {AirSim}
\bibfield{author}{\bibinfo{person}{microsoft}.}
  \bibinfo{year}{2021}\natexlab{a}.
\newblock \bibinfo{title}{AirSim}.
\newblock \bibinfo{howpublished}{\url{https://bit.ly/3qREI8Q}}.
\newblock


\bibitem[microsoft(2021b)]%
        {VoTT}
\bibfield{author}{\bibinfo{person}{microsoft}.}
  \bibinfo{year}{2021}\natexlab{b}.
\newblock \bibinfo{title}{UVoTT}.
\newblock \bibinfo{howpublished}{\url{https://bit.ly/3rEWxJO}}.
\newblock


\bibitem[Misherghi and Su(2006)]%
        {misherghi2006hdd}
\bibfield{author}{\bibinfo{person}{G. Misherghi} {and} \bibinfo{person}{Z.
  Su}.} \bibinfo{year}{2006}\natexlab{}.
\newblock \showarticletitle{HDD: Hierarchical delta debugging}. In
  \bibinfo{booktitle}{\emph{ICSE}}. \bibinfo{pages}{142--151}.
\newblock


\bibitem[Mullins et~al\mbox{.}(2018)]%
        {mullins2018adaptive}
\bibfield{author}{\bibinfo{person}{Galen~E Mullins}, \bibinfo{person}{Paul~G
  Stankiewicz}, \bibinfo{person}{R~Chad Hawthorne}, {and}
  \bibinfo{person}{Satyandra~K Gupta}.} \bibinfo{year}{2018}\natexlab{}.
\newblock \showarticletitle{Adaptive generation of challenging scenarios for
  testing and evaluation of autonomous vehicles}.
\newblock \bibinfo{journal}{\emph{Journal of Systems and Software}}
  \bibinfo{volume}{137} (\bibinfo{year}{2018}), \bibinfo{pages}{197--215}.
\newblock


\bibitem[Northcutt et~al\mbox{.}(2021)]%
        {northcutt2021pervasive}
\bibfield{author}{\bibinfo{person}{C. Northcutt}, \bibinfo{person}{A. Athalye},
  {and} \bibinfo{person}{J. Mueller}.} \bibinfo{year}{2021}\natexlab{}.
\newblock \showarticletitle{Pervasive label errors in test sets destabilize
  machine learning benchmarks}.
\newblock \bibinfo{journal}{\emph{arXiv preprint arXiv:2103.14749}}
  (\bibinfo{year}{2021}).
\newblock


\bibitem[of~Motor~Vehicles(2019)]%
        {CADMV}
\bibfield{author}{\bibinfo{person}{California~Department of Motor~Vehicles}.}
  \bibinfo{year}{2019}\natexlab{}.
\newblock \bibinfo{title}{Autonomous Vehicle Collision Reports - California
  DMV}.
\newblock \bibinfo{howpublished}{\url{https://bit.ly/3cPUcGC}}.
\newblock


\bibitem[of~Motor~Vehicles(2021)]%
        {Disengagement}
\bibfield{author}{\bibinfo{person}{California~Department of Motor~Vehicles}.}
  \bibinfo{year}{2021}\natexlab{}.
\newblock \bibinfo{title}{Disengagement Report}.
\newblock \bibinfo{howpublished}{\url{https://bit.ly/2NmXA1c}}.
\newblock


\bibitem[Paolacci et~al\mbox{.}(2014)]%
        {paolacci2014inside}
\bibfield{author}{\bibinfo{person}{G. Paolacci}, \bibinfo{person}{Ch}, {and}
  \bibinfo{person}{J. ler}.} \bibinfo{year}{2014}\natexlab{}.
\newblock \showarticletitle{Inside the Turk: Understanding Mechanical Turk as a
  participant pool}.
\newblock \bibinfo{journal}{\emph{Current directions in psychological science}}
  \bibinfo{volume}{23}, \bibinfo{number}{3} (\bibinfo{year}{2014}),
  \bibinfo{pages}{184--188}.
\newblock


\bibitem[Peng et~al\mbox{.}(2020)]%
        {peng2020first}
\bibfield{author}{\bibinfo{person}{Z. Peng}, \bibinfo{person}{J. Yang},
  \bibinfo{person}{T.~H. Chen}, {and} \bibinfo{person}{L. Ma}.}
  \bibinfo{year}{2020}\natexlab{}.
\newblock \showarticletitle{A first look at the integration of machine learning
  models in complex autonomous driving systems: a case study on Apollo}. In
  \bibinfo{booktitle}{\emph{FSE}}. \bibinfo{publisher}{{ACM}},
  \bibinfo{pages}{1240--1250}.
\newblock


\bibitem[Queiroz et~al\mbox{.}(2019)]%
        {queiroz2019geoscenario}
\bibfield{author}{\bibinfo{person}{R. Queiroz}, \bibinfo{person}{T. Berger},
  {and} \bibinfo{person}{K. Czarnecki}.} \bibinfo{year}{2019}\natexlab{}.
\newblock \showarticletitle{GeoScenario: An open DSL for autonomous driving
  scenario representation}. In \bibinfo{booktitle}{\emph{IV}}. IEEE,
  \bibinfo{pages}{287--294}.
\newblock


\bibitem[Rahm and Do(2000)]%
        {rahm2000data}
\bibfield{author}{\bibinfo{person}{E. Rahm} {and} \bibinfo{person}{H. Do}.}
  \bibinfo{year}{2000}\natexlab{}.
\newblock \showarticletitle{Data cleaning: Problems and current approaches}.
\newblock \bibinfo{journal}{\emph{IEEE Data Eng. Bull.}} \bibinfo{volume}{23},
  \bibinfo{number}{4} (\bibinfo{year}{2000}), \bibinfo{pages}{3--13}.
\newblock


\bibitem[Roh et~al\mbox{.}(2019)]%
        {roh2019survey}
\bibfield{author}{\bibinfo{person}{Y. Roh}, \bibinfo{person}{G. Heo}, {and}
  \bibinfo{person}{S. Whang}.} \bibinfo{year}{2019}\natexlab{}.
\newblock \showarticletitle{A survey on data collection for machine learning: a
  big data-ai integration perspective}.
\newblock \bibinfo{journal}{\emph{IEEE Transactions on Knowledge and Data
  Engineering}} (\bibinfo{year}{2019}).
\newblock


\bibitem[Salay et~al\mbox{.}(2019)]%
        {salay2019safety}
\bibfield{author}{\bibinfo{person}{Rick Salay}, \bibinfo{person}{Matt Angus},
  {and} \bibinfo{person}{Krzysztof Czarnecki}.}
  \bibinfo{year}{2019}\natexlab{}.
\newblock \showarticletitle{A safety analysis method for perceptual components
  in automated driving}. In \bibinfo{booktitle}{\emph{ISSRE}}. IEEE,
  \bibinfo{pages}{24--34}.
\newblock


\bibitem[Sharma et~al\mbox{.}(2019)]%
        {sharma2019labelbox}
\bibfield{author}{\bibinfo{person}{M. Sharma}, \bibinfo{person}{D. Rasmuson},
  \bibinfo{person}{B. Rieger}, \bibinfo{person}{D. Kjelkerud}, {et~al\mbox{.}}}
  \bibinfo{year}{2019}\natexlab{}.
\newblock \showarticletitle{Labelbox: The best way to create and manage
  training data. software, LabelBox}.
\newblock \bibinfo{journal}{\emph{Inc, https://bit.ly/2TBLzYW}}
  (\bibinfo{year}{2019}).
\newblock


\bibitem[Shull et~al\mbox{.}(2008)]%
        {DBLP:books/sp/08/SSS2008}
\bibfield{editor}{\bibinfo{person}{Forrest Shull}, \bibinfo{person}{Janice
  Singer}, {and} \bibinfo{person}{Dag I.~K. Sj{\o}berg}} (Eds.).
  \bibinfo{year}{2008}\natexlab{}.
\newblock \bibinfo{booktitle}{\emph{Guide to Advanced Empirical Software
  Engineering}}.
\newblock \bibinfo{publisher}{Springer}.
\newblock
\showISBNx{9781848000438}
\urldef\tempurl%
\url{https://doi.org/10.1007/978-1-84800-044-5}
\showDOI{\tempurl}


\bibitem[Sun et~al\mbox{.}(2018)]%
        {sun2018perses}
\bibfield{author}{\bibinfo{person}{C. Sun}, \bibinfo{person}{Y. Li},
  \bibinfo{person}{Q. Zhang}, \bibinfo{person}{T. Gu}, {and}
  \bibinfo{person}{Z. Su}.} \bibinfo{year}{2018}\natexlab{}.
\newblock \showarticletitle{Perses: Syntax-guided program reduction}. In
  \bibinfo{booktitle}{\emph{ICSE}}. \bibinfo{pages}{361--371}.
\newblock


\bibitem[Sun et~al\mbox{.}(2020)]%
        {sun2020scalability}
\bibfield{author}{\bibinfo{person}{P. Sun}, \bibinfo{person}{H. Kretzschmar},
  \bibinfo{person}{X. Dotiwalla}, \bibinfo{person}{A. Chouard},
  \bibinfo{person}{V. Patnaik}, \bibinfo{person}{P. Tsui}, \bibinfo{person}{J.
  Guo}, \bibinfo{person}{Y. Zhou}, \bibinfo{person}{Y. Chai},
  \bibinfo{person}{B. Caine}, {et~al\mbox{.}}} \bibinfo{year}{2020}\natexlab{}.
\newblock \showarticletitle{Scalability in perception for autonomous driving:
  Waymo open dataset}. In \bibinfo{booktitle}{\emph{CVPR}}.
  \bibinfo{pages}{2446--2454}.
\newblock


\bibitem[Sun et~al\mbox{.}(2017)]%
        {sun2017empirical}
\bibfield{author}{\bibinfo{person}{X. Sun}, \bibinfo{person}{T. Zhou},
  \bibinfo{person}{G. Li}, \bibinfo{person}{J. Hu}, \bibinfo{person}{H. Yang},
  {and} \bibinfo{person}{B. Li}.} \bibinfo{year}{2017}\natexlab{}.
\newblock \showarticletitle{An empirical study on real bugs for machine
  learning programs}. In \bibinfo{booktitle}{\emph{Asia-Pacific Software
  Engineering Conference}}. IEEE, \bibinfo{pages}{348--357}.
\newblock


\bibitem[Tang et~al\mbox{.}(2021)]%
        {tang2021systematic}
\bibfield{author}{\bibinfo{person}{Yun Tang}, \bibinfo{person}{Yuan Zhou},
  \bibinfo{person}{Tianwei Zhang}, \bibinfo{person}{Fenghua Wu},
  \bibinfo{person}{Yang Liu}, {and} \bibinfo{person}{Gang Wang}.}
  \bibinfo{year}{2021}\natexlab{}.
\newblock \showarticletitle{Systematic testing of autonomous driving systems
  using map topology-based scenario classification}. In
  \bibinfo{booktitle}{\emph{2021 36th IEEE/ACM International Conference on
  Automated Software Engineering (ASE)}}. IEEE, \bibinfo{pages}{1342--1346}.
\newblock


\bibitem[Tao et~al\mbox{.}(2019)]%
        {tao2019industrial}
\bibfield{author}{\bibinfo{person}{Jianbo Tao}, \bibinfo{person}{Yihao Li},
  \bibinfo{person}{Franz Wotawa}, \bibinfo{person}{Hermann Felbinger}, {and}
  \bibinfo{person}{Mihai Nica}.} \bibinfo{year}{2019}\natexlab{}.
\newblock \showarticletitle{On the industrial application of combinatorial
  testing for autonomous driving functions}. In \bibinfo{booktitle}{\emph{2019
  IEEE International Conference on Software Testing, Verification and
  Validation Workshops (ICSTW)}}. IEEE, \bibinfo{pages}{234--240}.
\newblock


\bibitem[Thung et~al\mbox{.}(2012)]%
        {thung2012empirical}
\bibfield{author}{\bibinfo{person}{F. Thung}, \bibinfo{person}{S. Wang},
  \bibinfo{person}{D. Lo}, {and} \bibinfo{person}{L. Jiang}.}
  \bibinfo{year}{2012}\natexlab{}.
\newblock \showarticletitle{An empirical study of bugs in machine learning
  systems}. In \bibinfo{booktitle}{\emph{ISSRE}}. IEEE,
  \bibinfo{pages}{271--280}.
\newblock


\bibitem[Tian et~al\mbox{.}(2018)]%
        {tian2018deeptest}
\bibfield{author}{\bibinfo{person}{Yuchi Tian}, \bibinfo{person}{Kexin Pei},
  \bibinfo{person}{Suman Jana}, {and} \bibinfo{person}{Baishakhi Ray}.}
  \bibinfo{year}{2018}\natexlab{}.
\newblock \showarticletitle{Deeptest: Automated testing of
  deep-neural-network-driven autonomous cars}. In
  \bibinfo{booktitle}{\emph{Proceedings of the 40th international conference on
  software engineering}}. \bibinfo{pages}{303--314}.
\newblock


\bibitem[Valle(2021)]%
        {valle2021metamorphic}
\bibfield{author}{\bibinfo{person}{Pablo Valle}.}
  \bibinfo{year}{2021}\natexlab{}.
\newblock \showarticletitle{Metamorphic testing of autonomous vehicles: a case
  study on simulink}. In \bibinfo{booktitle}{\emph{2021 IEEE/ACM 43rd
  International Conference on Software Engineering: Companion Proceedings
  (ICSE-Companion)}}. IEEE, \bibinfo{pages}{105--107}.
\newblock


\bibitem[Viera et~al\mbox{.}(2005)]%
        {viera2005understanding}
\bibfield{author}{\bibinfo{person}{A. Viera}, \bibinfo{person}{J. Garrett},
  {et~al\mbox{.}}} \bibinfo{year}{2005}\natexlab{}.
\newblock \showarticletitle{Understanding interobserver agreement: the kappa
  statistic}.
\newblock \bibinfo{journal}{\emph{Fam med}} \bibinfo{volume}{37},
  \bibinfo{number}{5} (\bibinfo{year}{2005}), \bibinfo{pages}{360--363}.
\newblock


\bibitem[Wang et~al\mbox{.}(2019)]%
        {wang2019apolloscape}
\bibfield{author}{\bibinfo{person}{P. Wang}, \bibinfo{person}{X. Huang},
  \bibinfo{person}{X. Cheng}, \bibinfo{person}{D. Zhou}, \bibinfo{person}{Q.
  Geng}, {and} \bibinfo{person}{R. Yang}.} \bibinfo{year}{2019}\natexlab{}.
\newblock \showarticletitle{The apolloscape open dataset for autonomous driving
  and its application}.
\newblock \bibinfo{journal}{\emph{IEEE transactions on pattern analysis and
  machine intelligence}} (\bibinfo{year}{2019}).
\newblock


\bibitem[Wang et~al\mbox{.}(2021)]%
        {wang2021prioritizing}
\bibfield{author}{\bibinfo{person}{Z. Wang}, \bibinfo{person}{H. You},
  \bibinfo{person}{J. Chen}, \bibinfo{person}{Y. Zhang}, \bibinfo{person}{X.
  Dong}, {and} \bibinfo{person}{W. Zhang}.} \bibinfo{year}{2021}\natexlab{}.
\newblock \showarticletitle{Prioritizing Test Inputs for Deep Neural Networks
  via Mutation Analysis}. In \bibinfo{booktitle}{\emph{ICSE}}. IEEE,
  \bibinfo{pages}{397--409}.
\newblock


\bibitem[Wohlin et~al\mbox{.}(2012)]%
        {Experimentation}
\bibfield{author}{\bibinfo{person}{C. Wohlin}, \bibinfo{person}{P. Runeson},
  \bibinfo{person}{M. Hst}, \bibinfo{person}{M. Ohlsson}, \bibinfo{person}{B.
  Regnell}, {and} \bibinfo{person}{A. Wessln}.}
  \bibinfo{year}{2012}\natexlab{}.
\newblock \bibinfo{booktitle}{\emph{Experimentation in Software Engineering}}.
\newblock \bibinfo{publisher}{Springer Publishing Company, Incorporated}.
\newblock
\showISBNx{3642290434}


\bibitem[Wolschke et~al\mbox{.}(2017)]%
        {wolschke2017observation}
\bibfield{author}{\bibinfo{person}{Christian Wolschke}, \bibinfo{person}{Thomas
  Kuhn}, \bibinfo{person}{Dieter Rombach}, {and} \bibinfo{person}{Peter
  Liggesmeyer}.} \bibinfo{year}{2017}\natexlab{}.
\newblock \showarticletitle{Observation based creation of minimal test suites
  for autonomous vehicles}. In \bibinfo{booktitle}{\emph{2017 IEEE
  International symposium on software reliability engineering workshops
  (ISSREW)}}. IEEE, \bibinfo{pages}{294--301}.
\newblock


\bibitem[Wu et~al\mbox{.}(2018)]%
        {wu2018tagging}
\bibfield{author}{\bibinfo{person}{B. Wu}, \bibinfo{person}{W. Chen},
  \bibinfo{person}{P. Sun}, \bibinfo{person}{W. Liu}, \bibinfo{person}{B.
  Ghanem}, {and} \bibinfo{person}{S. Lyu}.} \bibinfo{year}{2018}\natexlab{}.
\newblock \showarticletitle{Tagging like humans: Diverse and distinct image
  annotation}. In \bibinfo{booktitle}{\emph{CVPR}}.
  \bibinfo{pages}{7967--7975}.
\newblock


\bibitem[Xiang et~al\mbox{.}(2018)]%
        {xiang2018verification}
\bibfield{author}{\bibinfo{person}{W. Xiang}, \bibinfo{person}{P. Musau},
  \bibinfo{person}{A. Wild}, \bibinfo{person}{D.~M. Lopez}, \bibinfo{person}{N.
  Hamilton}, \bibinfo{person}{X. Yang}, \bibinfo{person}{J. Rosenfeld}, {and}
  \bibinfo{person}{T. Johnson}.} \bibinfo{year}{2018}\natexlab{}.
\newblock \showarticletitle{Verification for machine learning, autonomy, and
  neural networks survey}.
\newblock \bibinfo{journal}{\emph{arXiv preprint arXiv:1810.01989}}
  (\bibinfo{year}{2018}).
\newblock


\bibitem[Zeller et~al\mbox{.}(2002)]%
        {zeller2002simplifying}
\bibfield{author}{\bibinfo{person}{A. Zeller}, \bibinfo{person}{Hildebr}, {and}
  \bibinfo{person}{R. t}.} \bibinfo{year}{2002}\natexlab{}.
\newblock \showarticletitle{Simplifying and isolating failure-inducing input}.
\newblock \bibinfo{journal}{\emph{IEEE Transactions on Software Engineering}}
  \bibinfo{volume}{28}, \bibinfo{number}{2} (\bibinfo{year}{2002}),
  \bibinfo{pages}{183--200}.
\newblock


\bibitem[Zhang et~al\mbox{.}(2020b)]%
        {zhang2020machine}
\bibfield{author}{\bibinfo{person}{J. Zhang}, \bibinfo{person}{M. Harman},
  \bibinfo{person}{L. Ma}, {and} \bibinfo{person}{Y. Liu}.}
  \bibinfo{year}{2020}\natexlab{b}.
\newblock \showarticletitle{Machine learning testing: Survey, landscapes and
  horizons}.
\newblock \bibinfo{journal}{\emph{IEEE Transactions on Software Engineering}}
  (\bibinfo{year}{2020}).
\newblock


\bibitem[Zhang et~al\mbox{.}(2018b)]%
        {zhang2018deeproad}
\bibfield{author}{\bibinfo{person}{Mengshi Zhang}, \bibinfo{person}{Yuqun
  Zhang}, \bibinfo{person}{Lingming Zhang}, \bibinfo{person}{Cong Liu}, {and}
  \bibinfo{person}{Sarfraz Khurshid}.} \bibinfo{year}{2018}\natexlab{b}.
\newblock \showarticletitle{Deeproad: Gan-based metamorphic testing and input
  validation framework for autonomous driving systems}. In
  \bibinfo{booktitle}{\emph{2018 33rd IEEE/ACM International Conference on
  Automated Software Engineering (ASE)}}. IEEE, \bibinfo{pages}{132--142}.
\newblock


\bibitem[Zhang et~al\mbox{.}(2019)]%
        {zhang2019empirical}
\bibfield{author}{\bibinfo{person}{T. Zhang}, \bibinfo{person}{C. Gao},
  \bibinfo{person}{L. Ma}, \bibinfo{person}{M. Lyu}, {and} \bibinfo{person}{M.
  Kim}.} \bibinfo{year}{2019}\natexlab{}.
\newblock \showarticletitle{An empirical study of common challenges in
  developing deep learning applications}. In \bibinfo{booktitle}{\emph{ISSRE}}.
  IEEE, \bibinfo{pages}{104--115}.
\newblock


\bibitem[Zhang et~al\mbox{.}(2020a)]%
        {9282713}
\bibfield{author}{\bibinfo{person}{Xudong Zhang}, \bibinfo{person}{Yan Cai},
  {and} \bibinfo{person}{Zijiang Yang}.} \bibinfo{year}{2020}\natexlab{a}.
\newblock \showarticletitle{A Study on Testing Autonomous Driving Systems}. In
  \bibinfo{booktitle}{\emph{2020 IEEE 20th International Conference on Software
  Quality, Reliability and Security Companion (QRS-C)}}.
  \bibinfo{pages}{241--244}.
\newblock
\urldef\tempurl%
\url{https://doi.org/10.1109/QRS-C51114.2020.00048}
\showDOI{\tempurl}


\bibitem[Zhang et~al\mbox{.}(2018a)]%
        {zhang2018empirical}
\bibfield{author}{\bibinfo{person}{Y. Zhang}, \bibinfo{person}{Y. Chen},
  \bibinfo{person}{S. Cheung}, \bibinfo{person}{Y. Xiong}, {and}
  \bibinfo{person}{L. Zhang}.} \bibinfo{year}{2018}\natexlab{a}.
\newblock \showarticletitle{An empirical study on TensorFlow program bugs}. In
  \bibinfo{booktitle}{\emph{ISSTA}}. \bibinfo{pages}{129--140}.
\newblock


\bibitem[Zhao et~al\mbox{.}(2019)]%
        {zhao2019assessing}
\bibfield{author}{\bibinfo{person}{Xingyu Zhao}, \bibinfo{person}{Valentin
  Robu}, \bibinfo{person}{David Flynn}, \bibinfo{person}{Kizito Salako}, {and}
  \bibinfo{person}{Lorenzo Strigini}.} \bibinfo{year}{2019}\natexlab{}.
\newblock \showarticletitle{Assessing the safety and reliability of autonomous
  vehicles from road testing}. In \bibinfo{booktitle}{\emph{ISSRE}}. IEEE,
  \bibinfo{pages}{13--23}.
\newblock


\bibitem[Zhou et~al\mbox{.}(2020)]%
        {zhou2020deepbillboard}
\bibfield{author}{\bibinfo{person}{Husheng Zhou}, \bibinfo{person}{Wei Li},
  \bibinfo{person}{Zelun Kong}, \bibinfo{person}{Junfeng Guo},
  \bibinfo{person}{Yuqun Zhang}, \bibinfo{person}{Bei Yu},
  \bibinfo{person}{Lingming Zhang}, {and} \bibinfo{person}{Cong Liu}.}
  \bibinfo{year}{2020}\natexlab{}.
\newblock \showarticletitle{Deepbillboard: Systematic physical-world testing of
  autonomous driving systems}. In \bibinfo{booktitle}{\emph{ICSE}}. IEEE,
  \bibinfo{pages}{347--358}.
\newblock


\end{thebibliography}

\end{document}